\documentclass{article}
\usepackage[utf8]{inputenc}
\usepackage{geometry}
\usepackage{graphicx}
\usepackage{mathtools}
\usepackage{multirow}
\usepackage{makecell}

\title{A Time To Event Framework For Multi-touch Attribution}
\author{Dinah Shender\thanks{Google}, Ali Nasiri Amini\footnotemark[1], Xinlong Bao\footnotemark[1], Mert Dikmen\footnotemark[1], Amy Richardson\thanks{At Google when this work was done},\\Jing Wang\footnotemark[1]}
\date{}

\newcommand{\tstar}{t^{*}}

\begin{document}
\maketitle

\begin{abstract}
    Multi-touch attribution (MTA) estimates the relative contributions of the multiple ads a user may see prior to any observed conversions. Increasingly, advertisers also want to base budget and bidding decisions on these attributions, spending more on ads that drive more conversions. We describe two requirements for an MTA system to be suitable for this application: First, it must be able to handle continuously updated and incomplete data. Second, it must be sufficiently flexible to capture that an ad’s effect will change over time. We describe an MTA system, consisting of a model for user conversion behavior and a credit assignment algorithm, that satisfies these requirements. Our model for user conversion behavior treats conversions as occurrences in an inhomogeneous Poisson process, while our attribution algorithm is based on iteratively removing the last ad in the path.
\end{abstract}

\section{Introduction}
One of the promises of online advertising has been the ability to tie together ad views or clicks with actual outcomes (e.g. purchases, website visits, etc), also known as conversions. This gives advertisers insight into both the effectiveness of their ads overall, and the relative effectiveness of different types of ads. Multi-touch attribution (MTA) achieves this second goal by estimating the relative contributions of the multiple ads a user may see prior to any observed conversions. With the right data, MTA can quantify which ads contributed to which types of conversions.

Increasingly, advertisers also want to base budget and bidding decisions on these attributions, increasing spend on ads that drive more conversions. Bidding models are constantly updated with the latest data, in order to respond to changes in ad effectiveness, business fluctuations, etc. Incorporating MTA results into a bidding system requires an MTA system that is also capable of ingesting continuously updated data. This data is by its nature incomplete: we don’t know how many users who saw ads yesterday will ultimately convert, only how many have converted so far. Therefore our first requirement is that our MTA system be capable of handling incomplete data.

A second requirement is that an MTA system be sufficiently flexible to capture that an ad’s immediate effect on a user is likely to be different than its effect several weeks later. In practice, this tends to mean that conversions occurring soon after an ad are more likely to have been caused by the ad (and therefore the ad tends to receive more credit) than conversions occuring long after the ad.

To handle both requirements, we propose an MTA system that models how ads affect a user’s conversion rate over time. Ad credit can then be distributed according to how ads change the estimated conversion rate. We propose a modeling approach that treats conversions as occurrences in an inhomogeneous Poisson process, similar to many survival models used in epidemiology and biostatistics, also known as a time-to-event model. In the remainder of this paper, we describe our approach in detail, as well as how to use the resulting model for attribution. The time-to-event methodology inspires the name TEDDA (Time to Event Data Driven Attribution) for our modelling approach.

Our model can be used with either observational data or data from randomized experiments set up to measure the causal (or incremental) effects of ads. Both types of data can be useful, depending on the goals and context. Observational data can be interpreted in terms of the correlation between showing ads and conversions, while data from randomized experiments allows for a causal estimate of the number of conversions caused by ads. The latter is the gold standard, but these experiments are often difficult or impossible to run, and may not be available on an ongoing basis. In these cases, advertisers may prefer to do MTA on correlational data over not having any sort of MTA, particularly if they have evidence, perhaps from previous randomized experiments, that the relative credit assigned to the ad types of interest is unchanged after accounting for incrementality, even if the absolute credit differs. These are choices that the modeller must make based on their knowledge of the application area, media types, and brands involved. Rather than discussing the pros and cons of observational and experimental data, this paper will focus on the overall system that can be used to model either type of data.

The remainder of this paper is organized as follows: In Section \ref{sec:prev_work} we discuss previous work in this area. Section \ref{sec:attr_reqs} goes into further detail about the key issues an MTA system must solve. In Section \ref{sec:proposed_model} we present our system, describing both the model for conversion occurrences and the attribution credit assignment methodology. Section \ref{sec:eval} discusses how to evaluate the quality and accuracy of the system. Section \ref{sec:sims} discusses possible future improvements to this system.

\section{Previous Work}
\label{sec:prev_work}
Traditionally marketers have evaluated their advertising results using Media Mix Modeling (MMM), which typically fits time series models to a few years of aggregated conversion data in order to compare broad classes of ads (e.g. \cite{de2016effectiveness,kireyev2016display}). While this can help in allocating overall marketing spend between channels, it cannot provide the kind of granular cross- and within-channel insight into specific types of ads that MTA obtains by leveraging individual user paths.

This more recent line of work on modelling individual user paths can be divided into models that rely on the sequence of user events and those that incorporate the timestamps of these events. The former category includes Markov chain methods such as \cite{li2014attributing,anderl2014mapping}, which treat both conversions and each ad channel as states in a kth-order Markov chain and estimate the probability of a user moving from one state to the next. Attribution credit in these models is based on the change in conversion probabilities when a channel is removed from the chain. \cite{dalessandro2012causally} and \cite{shao2011data} are similar in depending only on the sequence of ad events, but they each fit logistic regression models for a binary conversion outcome. They then use Shapley values to distribute attribution credit, using the probability of a conversion as the value function in the Shapley algorithm.

There are also several previous papers falling into the latter category, i.e. models that incorporate event timestamps to fit continuous time models for user conversion activity. This includes \cite{du2019causally}, which fits a recurrent neural network with the entire path as an input to estimate the conversion probability each day. This is then used as input to the Shapley value algorithm to distribute attribution credit. \cite{lewis2018incrementality} uses exponentially decaying ad effects to build a model to estimate the incremental effect of ads using experimental data. They use an attribution methodology similar to the one we will describe to then bid optimally based on that model. Similar to our model, \cite{zhang2014multi} and \cite{xu2014path} treat conversions as occurrences in a Poisson process. Both assume an exponentially decaying ad effect.  \cite{zhang2014multi} considers a more restricted  setting where users can convert at most once, similar to traditional survival analysis, and then fits an additive model that assumes there are no interactions between ads. \cite{xu2014path} models both conversion events and ad events in different channels as a set of mutually exciting Poisson processes, allowing ad events to affect not just the conversion probability, but also the probability of future ads. However, they consider only the aggregate conversion credit over the data set, not the credit per ad.

\section{Attribution Requirements}
\label{sec:attr_reqs}
MTA provides critical insights about the relative value of ads, which should affect an advertiser's willingness to spend on different types of ads. In the context of digital advertising, where advertisers compete in an auction to determine whose ad is shown, the fastest way to incorporate this information is to allow it to affect the advertiser's bid. In order to do this, an advertiser must have access to an MTA system with nearly real-time (e.g. daily) updates. This allows the advertiser to bid in accordance with the attribution results without sacrificing their responsiveness to business fluctuations, changes in their ads' effectiveness, or other novel trends.

We propose two key requirements that this type of attribution system should satisfy, both related to how the system should consider the effects of time.

\subsection{Requirement \#1}
The system must be able to handle incomplete or censored data.

Real-time user path data is fundamentally incomplete: If an ad was shown this morning at 10am, and it's now 6pm and there was no conversion, that does not mean that this ad resulted in 0 conversions. Rather, it resulted in 0 conversions over the first 8 hours. We do not know what will happen over the next day or week. Instead of  treating this observation as a 0 or negative response, we should treat this observation as being right-censored.

\begin{figure}[h]
    \centering
    \includegraphics[width=\textwidth]{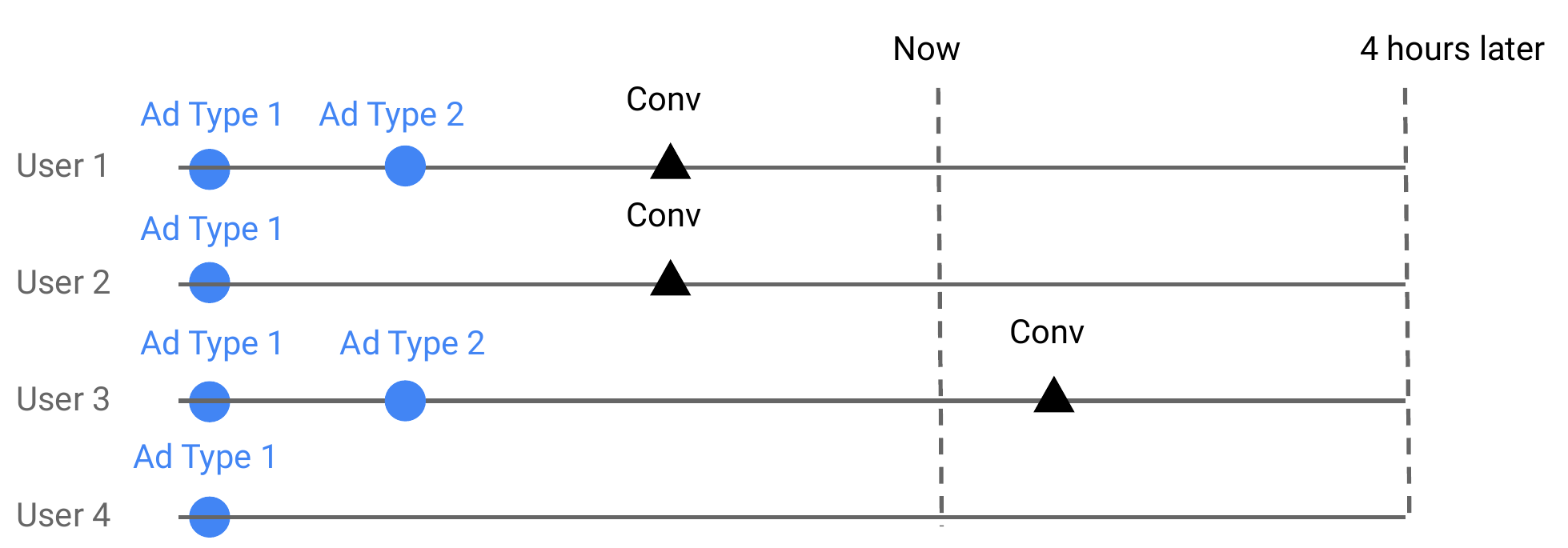}
    \caption{An example of the effects of incomplete or censored data}
    \label{fig:incomplete_info}
\end{figure}

To illustrate the importance of this principle, consider the toy example in Figure \ref{fig:incomplete_info}.

If our attribution system treats the data as being complete, so that the conversion outcome is binary for each user, with no censoring, and we run it now, it will likely find that having ad type 2 does not lead to additional conversions: paths with and without ad type 2 convert at equal rates\footnote{Note that this is not a causal claim about ad type 2 causing the conversions, but rather about the associations between ads and conversions. With observational data, we cannot know if ad type 2 causes more conversions or if users who see ad type 2 are simply more likely to convert anyways. If instead this is experimental data where users 2 and 4 didn't see ad type 2 because it was ablated or withheld, then we could turn this into a causal claim. The implications for our model are similar.}. However, in four hours, it would find that ad type 2 drives 50\% more conversions. Such strongly conflicting results can cause unstable and problematic behavior when used in bidding algorithms. To avoid this, our system needs to recognize that after 4 hours, we not only have new values for the response, but we also have more overall information about the rate of conversions because we have observed the user's post-ad paths for longer. In other words, we need to recognize that we have incomplete observations due to censoring.

This leads us to consider modeling the number of conversions over time, which would naturally capture that our estimates are more uncertain “now” than in 4 hours. The next requirement leads us further in this direction.

\subsection{Requirement \#2}
\label{subsec:req2}
An ad's contribution to any potential conversion should be allowed to differ depending on the time between the ad exposure and the conversion, typically decreasing as the time between the ad exposure and the conversion increases. Similarly, ad credit shouldn't just depend on the order of the ads in the path.

Many advertisers already recognize this in an informal way, setting “lookback windows” of $N=7$ or $30$ days and only distributing conversion credit to ads in the $N$ days before the conversion. However, an ad's influence likely decays more continuously, so that even within the lookback window, ads further in time from the conversion deserve less credit than those close to it, if all else is equal. While this implies that the order in which ads occur should affect the relative conversion credit that ads receive, the order is not sufficient: if two identical ads occur within hours of each other, we'd expect them to receive similar levels of credit for any subsequent conversions, whereas if they are separated by 30 days, we'd expect the later ad to receive more credit. How fast or steeply this decay occurs will depend on the individual advertiser and the type of ad, and should be learned from the data, but our model should be flexible enough to treat ads differently based on not just their order, but when they occurred.

\begin{figure}[h]
    \centering
    \includegraphics[width=\textwidth]{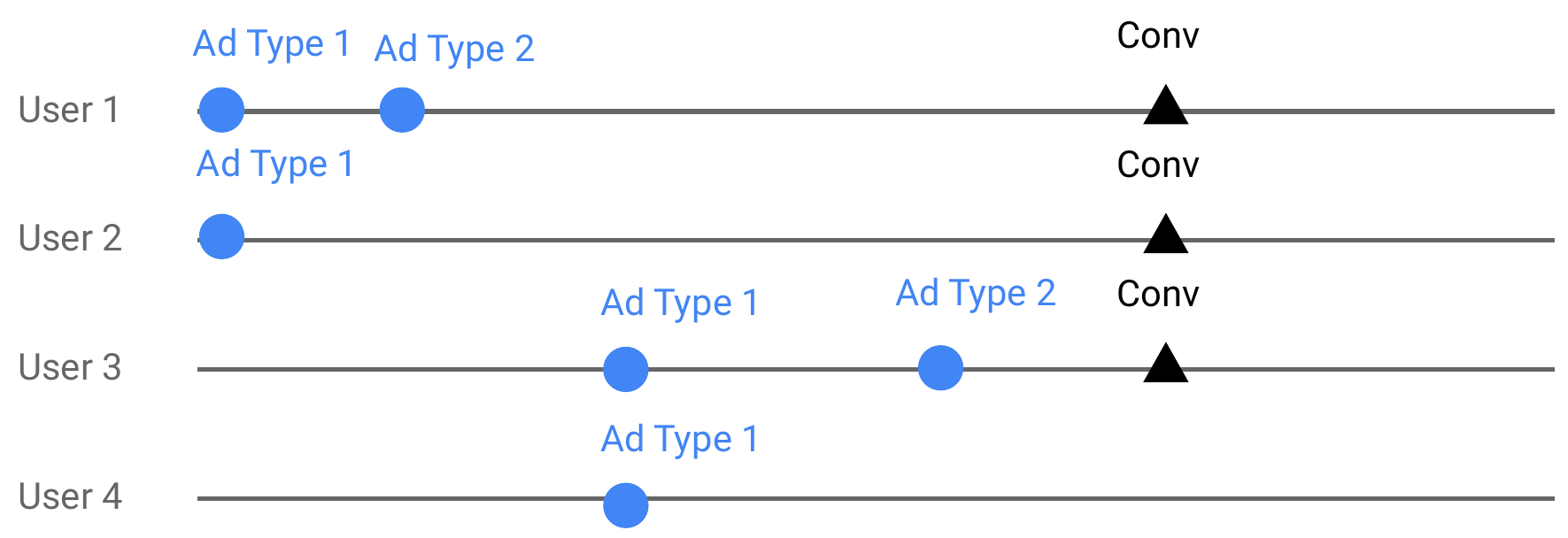}
    \caption{An example of the effects of the importance of considering event timestamps}
    \label{fig:timestamps}
\end{figure}

We illustrate this with another toy example in Figure \ref{fig:timestamps}. To distinguish this issue from Requirement \#1, let's suppose that the data is complete in this case, e.g. this data is from long enough ago that no further conversions are possible/expected.
If we look only at the order in which ads occur, we would learn that the presence of ad type 2 increases conversions by 50\%. But if we look at just users 1 and 2 and consider the timestamps, we will learn that the effect of ad type 2 decays fast enough that it deserves little credit for conversions far from it. However, looking at users 3 and 4, we will see that ad type 2 is actually responsible for all the conversions that occur soon after it is shown\footnote{As with the first example, these claims are all correlational. If instead Ad Type 2 were held back or ablated for users 2 and 4, then these claims would become causal, but as before, the implications for our model are similar.}. This is a simplified example, but it nevertheless illustrates that a model that looks only at the order of the ads and the binary conversion label can reach a very different conclusion compared to a model that takes into account the times at which ads and conversions occur. 

Together these two requirements imply that we should model not just the sequence of ad events and binary (or even integer) conversion outcomes for each path, but rather the times when conversions occur given when ads occur. A natural option for this type of modeling is survival analysis. In this framework, conversions are viewed as occurrences (or failures) of a Poisson process. The conversion intensity function, also known as the instantaneous occurrence rate or the hazard rate or just the intensity function, is allowed to vary with time, and in our case depends on the times at which a user has previously seen ads. Attribution credit is then based on the effect of an ad on a user’s instantaneous conversion rate at the time of the conversion.

Survival analysis techniques, including variations allowing for multiple occurrence, are widely used, particularly in biostatistics (e.g. \cite{cook2007statistical}). Indeed, in the case where a user can convert at most once, our model reduces to the classic Cox proportional hazards model with time-varying covariates \cite{andersen2012statistical}. Similar applications of survival analysis to multi-touch ad attribution have been considered by \cite{zhang2014multi,xu2014path} among others.  

In the next section we detail our model for conversions and how we use it for attribution.

\section{Proposed Model}
\label{sec:proposed_model}
Our attribution system has two parts: a model for users’ conversion behavior, and an attribution credit assignment algorithm that assigns credit in accordance with this model. We will focus the bulk of our exposition on the former. Given a model for user conversion behavior, there are many reasonable credit assignment algorithms. While we will discuss options and the algorithm that we use, the best choice depends on the goals of the system.

\subsection{Modeling User Conversion Behavior}
\label{subsection:user_model}
As discussed in the previous section, in order to capture how ad effects vary over time, as well as handle incomplete data, we will model conversions as a realization of a Poisson counting process with a time-varying intensity function, $\lambda(t)$. In particular, if we define $Y_i(t)$ as the number of occurrences (conversions) for user $i$ up until time $t$, then
\begin{equation}
    Y_i(t) - Y_i(s) \sim Poisson\left(\int_s^t \lambda(t)\mathrm{d}t\right)
\end{equation}
for any $0 \leq s \leq t$. We will use a log-linear model for the intensity, and allow it to depend on user features (e.g. the country the user is located in), the time since previous ads were seen, as well as other ad features (e.g., format of the ad). We will start with an overly simplified model for $\lambda(t)$ and gradually add these complexities. This model formulation treats data from randomized experiments in the same way as observational data, but with an additional feature for treatment group assignment. We will discuss specifics in a section towards the end. We will end the conversion modeling section with a brief discussion on options for estimating this model.

\subsubsection{No user features, no ad features, single ad per user}
\begin{figure}[h]
    \centering
    \includegraphics[width=\textwidth,trim=0.75in 1.75in 0.75in 1.75in, clip=true]{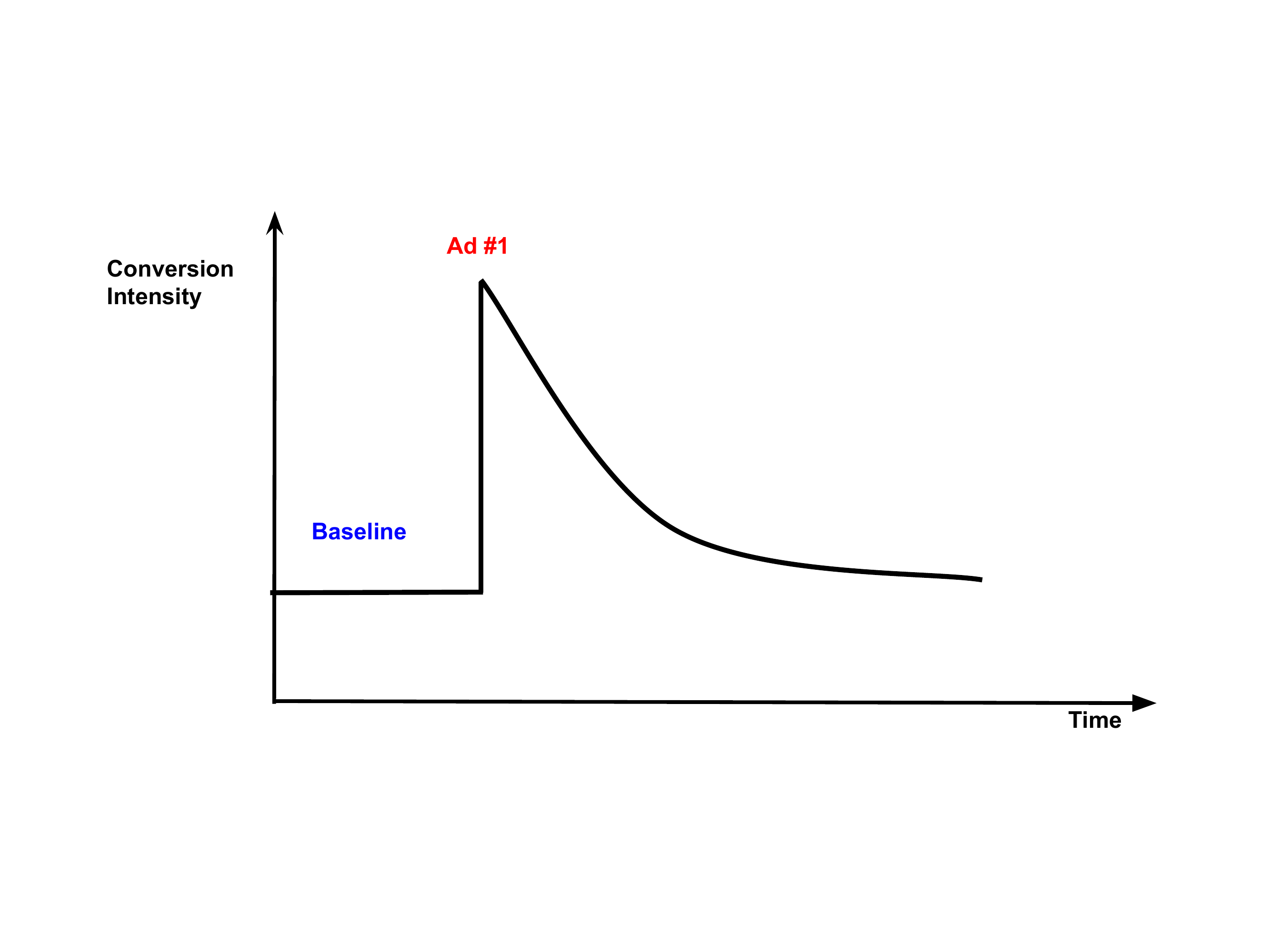}
    \caption{Conversion intensity over time for a user with a single ad event}
    \label{fig:single_ad_intensity}
\end{figure}

To start with, consider a model that has no user features, no ad features (other than time), and where we assume that each user sees at most one ad. This is an overly simplified view of the world, particularly the assumption of just a single ad per user, but it is useful for exposition. Suppose that the single ad occurs at $t_1$; in principle this should also be indexed by user, but we will drop the user subscripts for brevity. In this set-up, our model becomes
\begin{equation}
    \log(\lambda(t)) = \alpha_0 + f(t - t_1)
\end{equation}
$\alpha_0$ represents the $\log$ of the conversion rate before ads are shown, while $f(t-t_1)$ represents the effect of an ad on the user’s conversion rate. A user’s conversion intensity over time might then look as in Figure \ref{fig:single_ad_intensity}. The baseline rate before the ad is $\exp(\alpha_0)$, while after the ad the intensity is $\exp(\alpha_0) * \exp(f(t-t_1))$. Since this model is log-linear, ads have a multiplicative effect on users’ conversion rates.

In general, $f$ is a function of time that we would like to estimate. We restrict ourselves to functions such that $f(x)=0$ whenever $x\leq0$, i.e. there is no ad effect before the ad is actually seen. There are many ways to parameterize $f$ in order to do estimation; we will review a few concrete options, but many other parameterizations are also possible.

One option is to treat $f$ as a continuous function. Since we often expect the ad effect to decay rapidly, modeling $f$ as a mixture of exponentials is a natural choice. We can consider a basis of exponential decay functions $\{\exp(-\theta_l t)\}_l$, where the choices of $\theta_l$ determine the span of the basis and the speed of the decay. Then we can parameterize the intensity as
\begin{equation}
    \log(\lambda(t)) = \alpha_0 + \sum_{l=1}^L \beta_l \exp(-\theta_l(t-t_1))
\end{equation}
where $\beta_l$ are parameters to be estimated. The above equation is consistent with a proportional hazards model and essentially implies a doubly exponential decay for $\lambda(t)$. \cite{lewis2018incrementality} and \cite{zhang2014multi} proceed in a similar fashion, but set the right-hand side of the above equation equal to $\lambda(t)$ rather than $\log(\lambda(t))$, essentially assuming an additive hazards model.

Another option for estimating $f$ as a continuous function is to use splines. Letting ${b_1,\dots, b_L}$ be the functions in the spline basis, the model becomes
\begin{equation}
    \log(\lambda(t)) = \alpha_0 + \sum_{l=1}^L \beta_l b_l(t-t_1)
\end{equation}
We can then optimize for $\alpha_0$ and $\beta_1,\dots,\beta_L$, possibly with a regularization constraint or prior on the $\beta_l$ values. Splines are commonly used for approximating continuous functions. However, if the true intensity does decay very rapidly, then for a fixed basis size, an exponential function basis might perform better than a spline basis. We do not make a specific recommendation here; rather, this is an area for future study.

A third option might be to approximate $f$ as a step function, essentially estimating a separate ad effect for each step. For example, we might choose to estimate the jump in conversions in the first 24 hours, the next 24 hours, and the subsequent 28 days. Then if $t$ is measured in hours, and using $I$ to denote indicator functions, the log-intensity becomes:
\begin{equation}
\label{eq:curr_tedda_model}
    \log(\lambda(t)) = \alpha_0 + \beta_1 I\{t-t_1 \leq 24\} + \beta_2 I\{24<t-t_1 \leq 48\} +
                       \beta_3 I\{48<t-t_1 \leq 24*30\}
\end{equation}
Again, we can now optimize for $\alpha_0, \beta_1, \beta_2$, and $\beta_3$, perhaps placing a prior on the coefficients or applying other types of regularization. As we add additional features to the model or if we wish to pool data across advertisers, treating these coefficients as random effects may also be attractive.

\subsubsection{Ad Features}
Staying for now in the single ad setting, we can also consider how to add ad features, such as the format of the ad, whether it was shown on a mobile device, and so on, to the model. For example, perhaps certain ad formats or campaigns are associated with a larger change in conversions than others. To capture this, we want to allow the ad effect to vary depending on these features. One way to do this with $K$ total features is to write:
\begin{equation}
    \log(\lambda(t)) = \alpha_0 + f(t - t_1) + \sum_{k=1}^K g_k(t-t_1, x_{1k})
\end{equation}
Here $k$ indexes the features in our model and $x_{1k}$ is the value of the $k^{th}$ feature for the first ad. Like $f$, $g_k$ is a function of time to be estimated. One option is to constrain $g_k$ to be constant over time, so that changing the feature value simply shifts the intercept for $f$. Conceptually, this corresponds to a change in an ad’s initial effect, but not its decay rate. In the example above where $f$ is a linear combination of spline basis functions, we could take $g_k$ to be a linear combination of lower-order spline basis functions. This would change both the ad’s initial effect and the effect’s decay rate. While we would generally expect to choose $g_k$ to be simpler than $f$, even this is not required.

As written, if for each level of each feature $k$ we allow a non-zero value for $g_k$, the model is overparameterized. This can be handled in the usual ways, e.g.  setting $g_k=0$ for some reference level and interpreting $f$ as the ad effect for when all ad features are at their reference level, or by adding a constraint so that the $g_k$ average to 0 for each $k$ and interpreting $f$ as the average ad effect. Alternatively, if we are applying regularization when we fit the model, then that may be enough for all the parameters to be well-defined. Our model is flexible with respect to these choices and so we leave the notation general.

\subsubsection{Multiple Ads}
\begin{figure}[h]
    \centering
    \includegraphics[width=\textwidth,trim=0.5in 1.5in 0.75in 1in, clip=true]{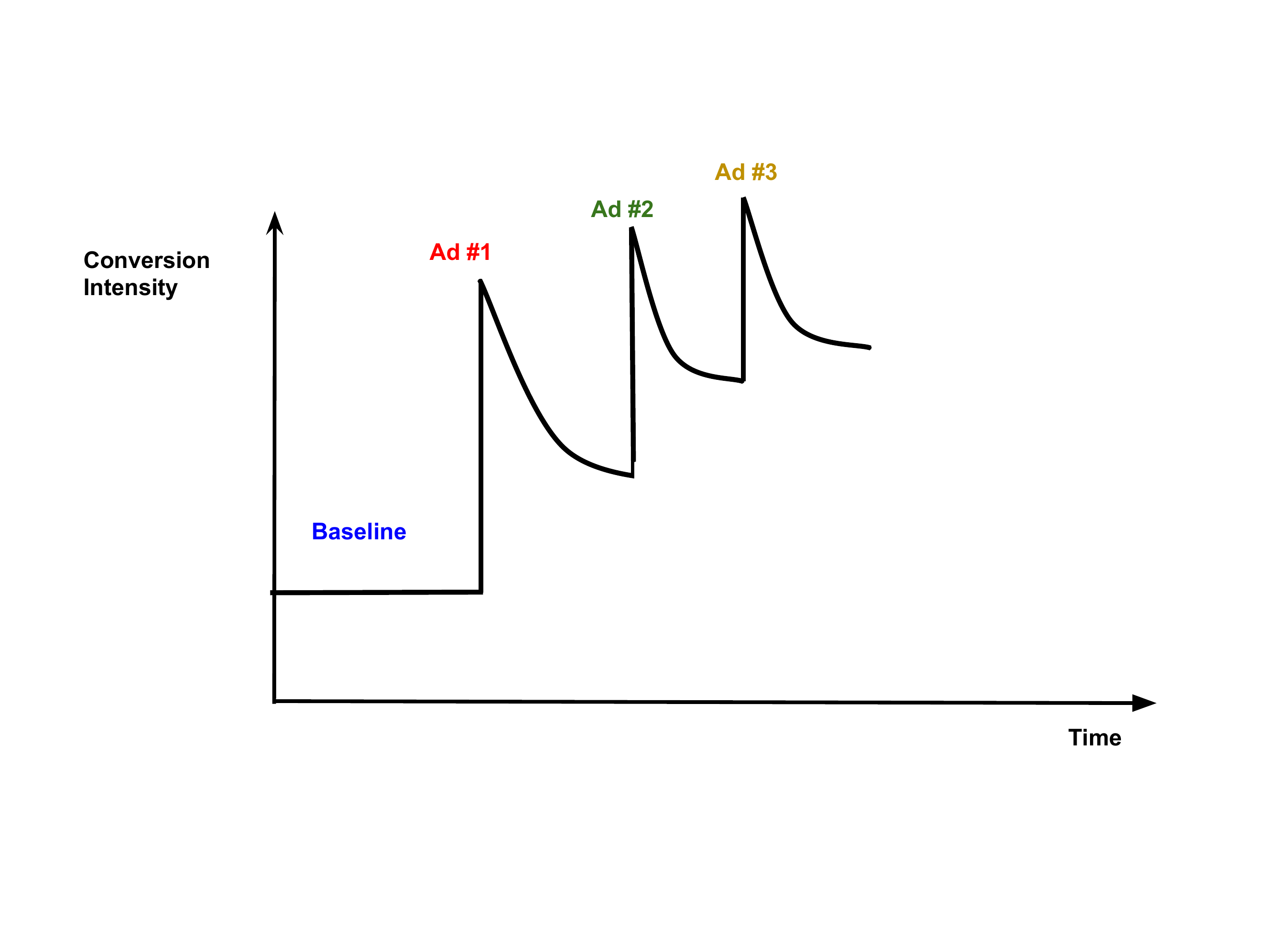}
    \caption{Conversion intensity over time for a user with multiple ad events}
    \label{fig:multiple_ad_intensity}
\end{figure}

The motivation for MTA is the case where users see multiple ads, making the models so far of expository, rather than practical, interest. Suppose now that a user sees multiple ads at times $t_1,\dots,t_J$. As a starting point, we consider the following model:
\begin{equation}
    \log(\lambda(t)) = \alpha_0 +\sum_j f(t - t_j) + \sum_{j,k} g_k(t-t_j, x_{jk})
\end{equation}
where $x_{jk}$ is the feature value for the $k^{th}$ feature for the $j^{th}$ ad. At a high level, this model says that ads with identical features have the same decay curve, but ads with different feature values may still have different decay curves. This is illustrated in Figure \ref{fig:multiple_ad_intensity}, which shows what an intensity curve for a user who sees three ads, each with different features values, might look like.

While this is not easily illustrated in a figure, notice that our model formulation also encodes that the effects of multiple ads are multiplicative (and therefore additive on a log-scale). This does not entirely match with our intuition: if showing a user an ad doubles their instantaneous conversion rate in the short-term, we wouldn’t expect showing them 50 ads within 10 minutes (perhaps with multiple ads per site) to increase their conversion rate by a factor of $2^{50}$. Simultaneously, it’s intuitive that there might be a synergy between ads, such that seeing ad $A$ might increase conversions by 2x, seeing ad $B$ might increase conversions by 1.5x, but that seeing both ads increases conversions by 5x rather than 3x. In other words, interactions between ads are possible and our model needs to be flexible enough for the user to fit interaction effects, should they choose to.

One way to accomplish this is to generalize our notation slightly and allow $x_{jk}$ to depend not only on ad $j$, but also on ads $j'< j$, as well as the subscript value $j$ itself. In other words, the features can depend not only on the current ad, but on previous ads and the ad index.

This allows us to add an effect for being the  $j^{th}$ ad. For example, assuming for simplicity that we have no other ad features, we could define a single ad feature $g_1(t-t_j, j) = f_j(t-t_j) - f(t-t_j)$, where $f_j$ is the effect function for the $j^{th}$ ad and $f$ is the “default” or average ad effect function. Then our model formulation is equivalent to letting
\begin{equation}
    \log(\lambda(t)) = \alpha_0 + \sum_j f_j(t - t_j)
\label{eq:model_ads_individually}
\end{equation}
i.e. a model where the ad effect differs for each ad.

Allowing the feature vector for an ad, $x_j=(x_{j1},\dots,x_{jK})$, to depend on previous ads also allows us to encode interactions between different ad formats or the timing of different ads. For example, we could add a feature $I(t_j-t_{j'} < \Delta  \text{ for } j'<j)$ as an indicator for whether there was a preceding ad in the previous $\Delta$ hours. This can be desirable if we only want to consider ad interactions if the two ads are within $\Delta$ hours of each other. Assuming for notational simplicity that this is the only feature we model, then we have
\begin{equation}
\label{eq:ad_interactions}
    \log(\lambda(t)) = \alpha_0 + \sum_j f(t - t_j)  +
                                \sum_j g_1(t-t_j) I(t_j-t_{j'} < \Delta \text{ for } j'<j)
\end{equation}
where, without loss of generality, we have set $I(t_j-t_{j'} < \Delta \text{ for } j'<j) = 0$ as the “default” or reference level.

The complexity of the model and any included interactions are necessarily application-dependent. In the case of an advertiser with a large amount of data and many ads shown per user, we may be able to make detailed estimates of the marginal effect of successive ads, as well as their interactions. On the other hand, when data is scarcer, it may be more practical to assume all ads have the same effect. The key point here is that this framework is flexible enough to allow for many different specifications of the factors that affect the conversion rate.

\subsubsection{User Features}
User features that affect conversion rates can be handled in one of two ways. If the feature only changes the user’s overall conversion rate, but not how they react to ads, then we can treat it as a shift in $\alpha_0$. Taking as an example the case of a model with a single user feature, a user’s bucketized age, we write
\begin{equation}
    \log(\lambda(t)) = \alpha_0 + \alpha_{\text{age bucket}} +
                                \sum_j f(t - t_j) + \sum_{j,k} g_k(t-t_j, x_{jk})
\end{equation}
If instead we believe that the user feature changes the ad effect, then we can incorporate it into the model in the same way as any other ad feature, $x_{jk}$, thinking of the feature as “age when this ad was shown.” Of course, unlike other ad features, which may vary amongst ads on the same path, this one will not, but that does not change how we write our model. As long as this feature varies across users, it will still be estimable.

\subsubsection{Experimental Data}
So far we have considered a general model, which is applicable to all data and which, absent additional assumptions or prior experimental data, estimates the correlational effect of ads on conversion intensity. However, as mentioned in the introduction, with experimental data we can measure the causal effect of ads on conversion rates. These can be incorporated into our model.

The design of experiments measuring the causal effects of ads is highly dependent on the details of the media type and ad serving environment. For our discussion,  we assume  a generic design in which some ads are shown (“exposed”), while others are withheld (“unexposed”), but where we still log when an advertiser’s ad would have been shown had we not withheld it.  We call the event where we receive a request for an ad a query event, and an event where the ad is actually returned an ad event. Thus when ads are shown, there are two simultaneous events, an ad event and a query event, while when ads are not shown there is a single query event.

Then consider the model
\begin{align}
\label{eq:incr_model}
    \log(\lambda(t)) = \alpha_0 + \alpha_{\text{age bucket}} &+ \sum_j f(t - t_j) I\{\text{ad $j$ shown}\} + \sum_{j,k} g_k(t-t_j, x_{jk}) I\{\text{ad $j$ shown}\} \nonumber \\
& {} + \sum_j m(t - t_j) +\sum_{j,k} n_k(t-t_j, x_{jk})
\end{align}
 $m_j(t-t_j) = m(t-t_j) + \sum_k n_k(t-t_j, x_{jk})$ represent the observed change in a user’s conversion rate (on a log scale) after the $j^{th}$ query. Note that as with the previous observational data, this query effect is not necessarily a causal effect: being targeted for an ad does not lead to you then making a purchase. If you use a search engine to search for “sneakers”, you’re more likely to buy sneakers after the query than a user who doesn’t do that search, but it probably wasn’t the query that caused that difference. Rather, you were interested in sneakers, and then you did the search.

\begin{figure}[h]
    \centering
    \includegraphics[width=\textwidth,trim=0.5in 1.5in 0.75in 1in, clip=true]{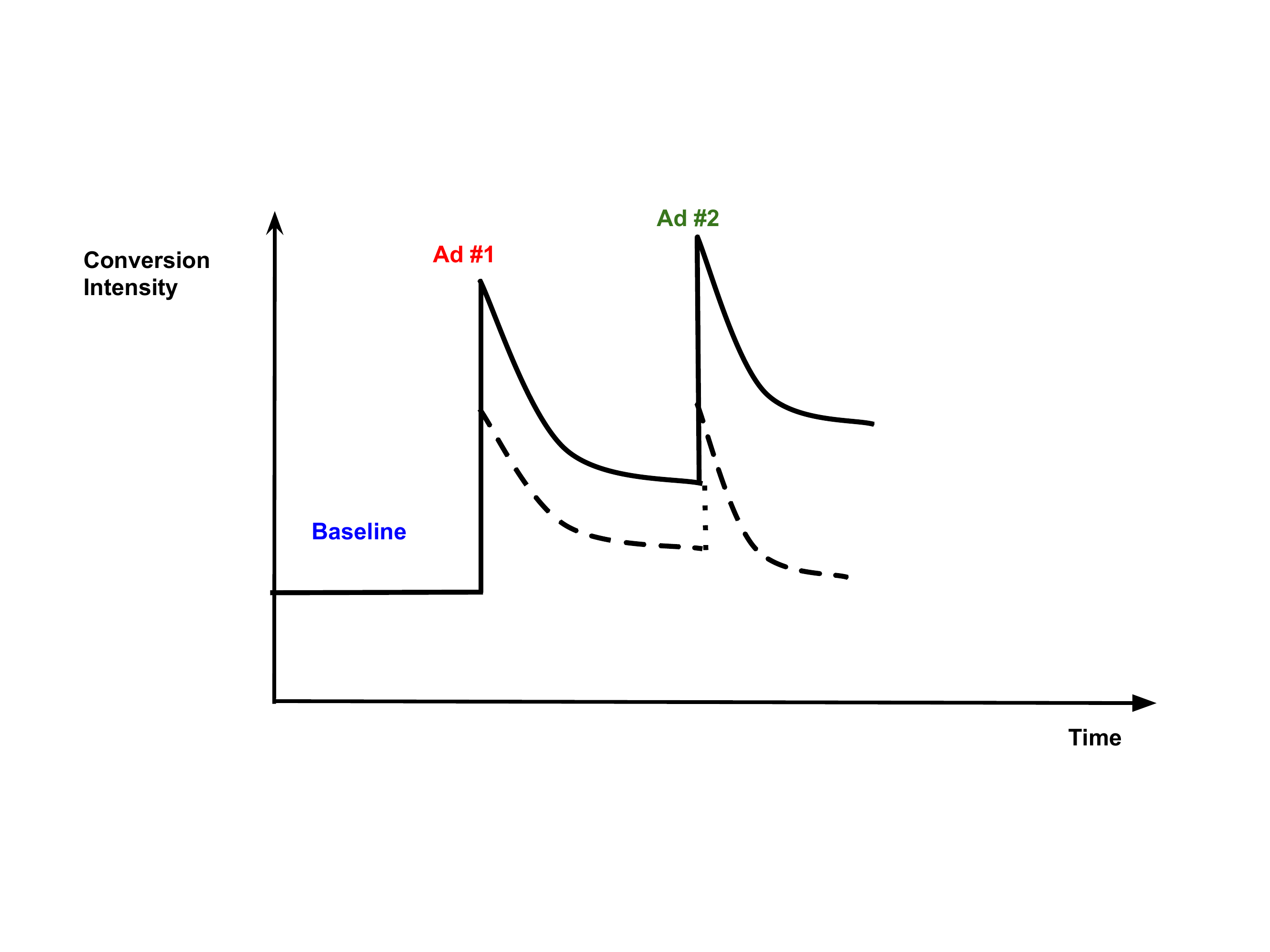}
    \caption{Conversion intensity over time for a user who saw ads (solid line) vs user with the same queries for whom ads were held back (dashed line).}
    \label{fig:incremental_intensity}
\end{figure}

The ad effect, separate from any query effects, for the $j^{th}$ ad is given by  $f_j(t-t_j) = f(t-t_j) + \sum_k g_k(t-t_j, x_{jk})$. This is the additional increase (on a log scale) in a user’s conversion intensity if the ad is actually shown. Figure \ref{fig:incremental_intensity} shows the intensity curves for a user who saw ads (solid line) and a user who had the same queries but for whom ads were withheld. Their difference would represent the ad effect.

Whether our experiment is such that $f_j$, the ad effect for a user’s $j^{th}$ query, is causally identifiable depends on the details of our experiment, as well as how ads are served in our setting. In the generic design described above, the overall causal effect of all the ads subject to ablation, i.e. the number of incremental conversions, is clearly identifiable. However, if there are multiple types of ads (e.g. comparing effects of first and second ad or search ads and display ads), then either changes in the experimental design (e.g. randomly ablating subsets of ads for each user) or additional assumptions (e.g. order of ads doesn’t matter) may be needed for all the effects to be causally identifiable. The details of this kind of experimental design are out of scope for this paper. However, in our experience, quite often some function of the $f_j$ has a valid causal estimator that can be derived from our model.

\subsubsection{Further Refinements}
So far the model has assumed that only ads, queries, and user features change the conversion rate. However, there are other factors which can also change conversion rates. In general, our model is flexible enough to accommodate many of these.

One common factor is seasonal effects, such as holidays, or more granular effects due to time of day or day of week. Modeling these can be as simple as adding indicators for each hour of day, or as complex as fitting additional spline functions to capture the pattern.

Another factor is that conversions themselves affect future rates of conversion. At one extreme are cases where only a single conversion is possible for each user, such as if the conversion represents signing up for a service or downloading an app. In those cases we can treat the model as a single-occurrence survival model and estimate it accordingly. In less extreme cases a conversion might decrease the likelihood of future conversions (e.g. someone buying a vacation package is probably less likely to buy a second one right away), but it can also increase the likelihood of a conversion (e.g. someone buying clothes from a retailer might be more likely to purchase from them again in future). Thus treating conversions themselves as events that can change the post-conversion intensity may improve the model. This is done by \cite{xu2014path}, which models both ad clicks and conversions as mutually exciting Poisson processes, where the intensity of each process depends on both itself and the other processes. While we may not want to model the ad clicks explicitly in our setting, we may consider a similar approach of letting the conversion intensity depend on the number or timing of past conversions.

\subsubsection{Estimation}
There are different ways to estimate the intensity function of a Poisson process. If $f, g_k$ are either piecewise constant or can be approximated as such, then by breaking each user’s path into intervals where the intensity, $\lambda(t)$, is constant, we can treat each interval as an observation in a Poisson regression problem. The number of conversions in the interval is then the response, and the length of the interval is the offset.

There is a large literature on fitting these types of large-scale regression problems, Poisson and otherwise. The details are outside the scope of this paper and we will not attempt to survey the existing technologies, but point to \cite{johnson2016scalable} as one example of a practical solution. There the authors place a prior on the parameters and use Bayesian machine learning methods to estimate the parameters and their prior variances, essentially treating the parameters as random effects.

If we do not want to assume that $f, g_k$ are (approximately) piecewise constant, then we can use the likelihood for an inhomogeneous Poisson process directly. Suppose we observe users for the time interval $[0, \tau]$. Let $\lambda_i(t)$ be the conversion intensity for user $i$ at time $t$, and let $T_{ij}, j=1,...,C_i$ be the conversion times for user $i$. The log-likelihood is
\begin{equation}
    \sum_{i=1}^N \left[ -\int_0^\tau \lambda_i(t) \mathrm{d}t + \sum_{j=1}^{C_i} \log( \lambda_i(T_{ij})) \right]
\end{equation}
See \cite{andersen2012statistical} for a detailed derivation. In the case where $f, g_k$ (and therefore $\lambda$) are piecewise constant, this reduces to the log-likelihood for the Poisson regression approach above. Without the piecewise constant assumption, one could instead try standard optimization techniques, such as gradient descent, to estimate the parameters for $f, g_k$.
\subsection{Credit Assignment Algorithm}
Even given a model for conversions, there are still many ways to distribute credit for the observed conversions to preceding ads. Different methodologies with different properties may be desirable depending on the context. The method we propose, which we call backwards elimination, differs from existing methods in how it distributes credit for conversions (or changes in conversion intensity) that only occur because users were shown multiple ads. Our algorithm tends to give this credit to later ads, while Shapley value-based methods, commonly used in the existing literature, divide this credit evenly amongst the ads. Both methods are reasonable, but they solve different problems, as we will discuss later.

We will first introduce backwards elimination in detail, including considerations for attribution with experimental data. We then examine more closely how backwards elimination distributes credit due to synergies between ads and compare this to Shapley values.

\subsubsection{Backwards Elimination}
\label{subsubsec:backwards_elimination}
\begin{figure}[h]
    \centering
    \includegraphics[width=\textwidth,trim=0.5in 1.5in 0.5in 1in, clip=true]{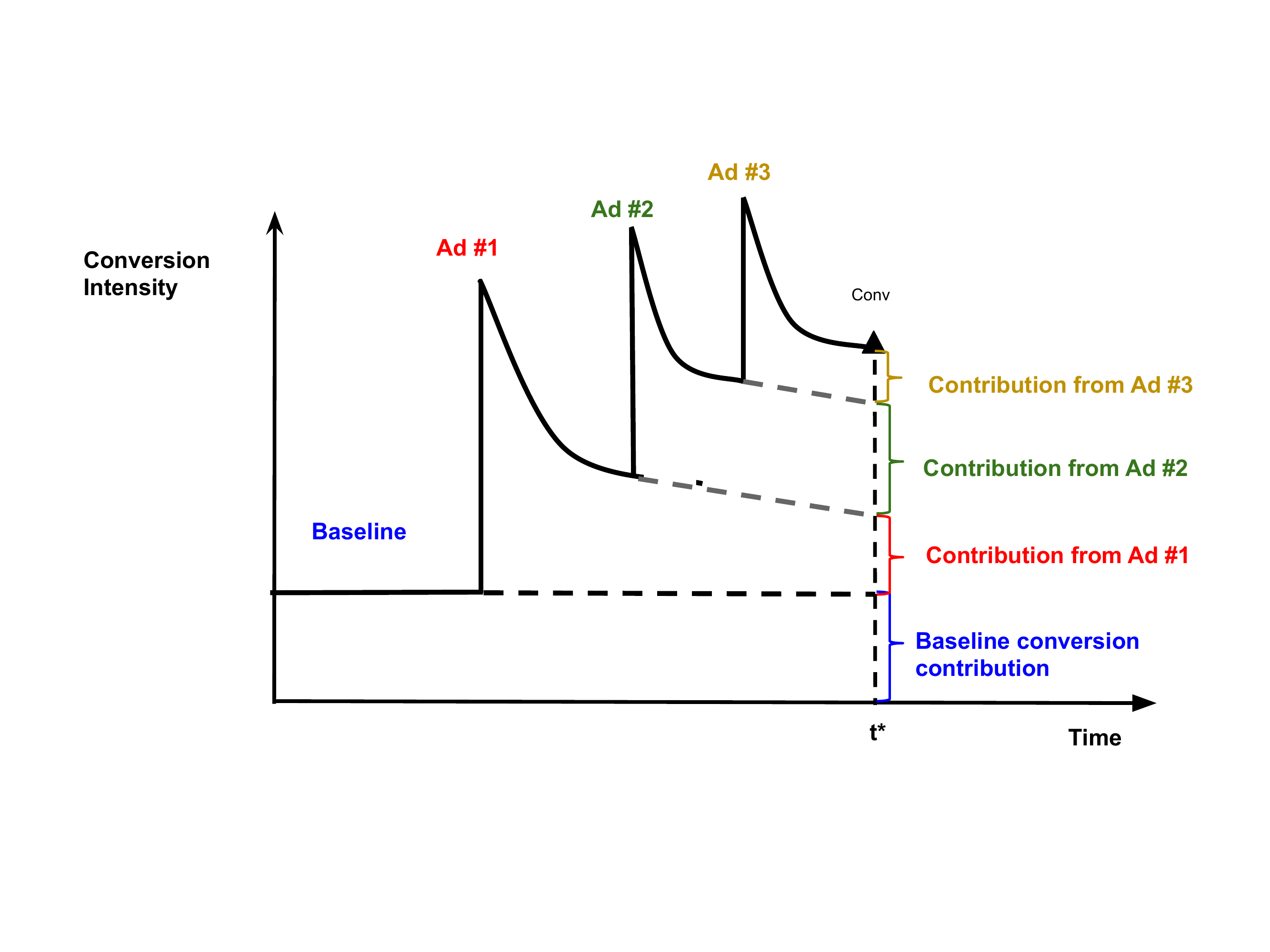}
    \caption{Attributed contribution of each ad to a user conversion occurring at $\tstar$.}
    \label{fig:attr_example}
\end{figure}

To illustrate the backwards elimination algorithm, consider a user path with three ads followed by a conversion at time $\tstar$, and whose estimated conversion intensity is plotted in Figure \ref{fig:attr_example}.

We define the contribution from the last ad before the conversion to be the difference in the estimated conversion intensity at $\tstar$ with all ads minus the estimated intensity at $\tstar$ if the last ad is dropped. The contribution from the second-to-last ad is the difference in conversion intensity at time $\tstar$ if the last ad is dropped minus the intensity if the last two ads are dropped. More generally, we proceed backwards through the path, removing an additional ad and attributing credit to the removed ad in proportion to the resulting change in intensity.

More formally, let $\mathcal{A}(n) = \{(X_j, t_j): j=1,...,n\}$ denote the first $n$ ads on user $i$’s path, where we have dropped the index $i$ for convenience, and where $X_j=(x_{j1},...,x_{jk},...,x_{jK})$ is the vector of ad features in our model. Without loss of generality, suppose that $t_n<\tstar<t_{n+1}$, i.e. that the conversion of interest occurs after the $n^{th}$ ad but before the $n+1^{st}$ ad. Let $\hat{\lambda}(t, \mathcal{A}(j))$ be the estimated conversion intensity at time $t$ for a user who sees the ads in $\mathcal{A}(j)$. Then the raw credit from our algorithm is
\begin{equation}
    RawCredit(j) = \hat{\lambda}(\tstar, \mathcal{A}(j)) - \hat{\lambda}(\tstar, \mathcal{A}(j-1))
\end{equation}
where we let $\mathcal{A}(0) = \emptyset$. We can also define the baseline credit as $RawCredit(baseline) = \hat{\lambda}(\tstar, \emptyset)$. Notice that the total raw credit given to ads equals $\hat{\lambda}(\tstar, \mathcal{A}(n)) - \hat{\lambda}(\tstar, \emptyset))$, which in turn equals the difference in the instantaneous conversion rate at $\tstar$ if all ads are dropped. This follows straightforwardly from the telescoping nature of the formula above.

As an example, suppose the estimated model for $\hat{\lambda}(t)$ is of the form
\begin{equation}
    \log(\hat{\lambda}(t)) = \hat{\alpha}_0 + \hat{\alpha}_{\text{age bucket}} + 
                             \sum_j \hat{f}(t - t_j)  + \sum_{j,k} \hat{g}_k(t-t_j, x_{jk})
\end{equation}
Then for the $l^{th}$ ad, we have 
\begin{equation}
    \log(\hat{\lambda}(\tstar, \mathcal{A}(l))) = \hat{\alpha}_0 + \hat{\alpha}_{\text{age bucket}} + 
                        \sum_{j=1}^l \hat{f}(\tstar - t_j)  + \sum_{j=1}^l \sum_k \hat{g}_k(\tstar-t_j, x_{jk}) 
\end{equation}
With this model for successive ads $\hat{\lambda}(\tstar, \mathcal{A}(l))$ will differ from $\hat{\lambda}(\tstar, \mathcal{A}(l-1))$ by a multiplicative factor of $\exp(\hat{f}(\tstar - t_l) + \sum_k \hat{g}_k(\tstar-t_l, x_{lk}))$ so that
\begin{align}
    RawCredit(l) &= \hat{\lambda}(\tstar, \mathcal{A}(l)) - \hat{\lambda}(\tstar, \mathcal{A}(l-1)) \nonumber \\
    &= \hat{\lambda}(\tstar, \mathcal{A}(l-1)) \left( \exp\left(\hat{f}(\tstar - t_l) + \sum_k \hat{g}_k(\tstar-t_l, x_{lk})\right) - 1\right)
\end{align}

There are two natural ways to normalize the raw credit. We can consider either 
\begin{equation}
    NormalizedCredit(j) = \frac{\hat{\lambda}(\tstar, \mathcal{A}(j)) - \hat{\lambda}(\tstar, \mathcal{A}(j-1))}{\hat{\lambda}(\tstar, \mathcal{A}(n))}
\end{equation}
or
\begin{equation}
    NonBaselineNormalizedCredit(j) = \frac{\hat{\lambda}(\tstar, \mathcal{A}(j)) - \hat{\lambda}(\tstar, \mathcal{A}(j-1))}{\hat{\lambda}(\tstar, \mathcal{A}(n)) - \hat{\lambda}(\tstar, \emptyset)}
\end{equation}
If we use $NormalizedCredit(j)$, and normalize $RawCredit(baseline)$ in a similar way, then the total ad credit plus the baseline credit will equal the number of conversions. If we instead use $NonBaselineNormalizedCredit(j)$, then the total ad credit will equal the total number of conversions.

It can be shown that, assuming our estimated intensity is correct, the total expected credit from all conversion occurrences in the path is $E[NormalizedCredit(j)] = \int \left( \hat{\lambda}(t, \mathcal{A}(j)) - \hat{\lambda}(t, \mathcal{A}(j-1)) \right) \mathrm{d}t$ or the difference in the expected number of conversions for a user seeing the ads in $\mathcal{A}(j)$ versus a user seeing the ads in $\mathcal{A}(j-1)$. See appendix \ref{sec:appendix} for details. Thus the expected ad credit for ad $j$ is the expected number of additional conversions gained when it was added to the end of the path, without considering gains due to combining ad $j$ with later ads.

One implication of this is that for a path with multiple ads, credit for “extra” conversions that occur because the ads are shown to the same user, rather than each ad being shown to different users, goes to the later ads, at least in expectation. In fact, this holds not just in expectation and is actually a property of the raw credit, not just the normalized credit. The details are in section \ref{subsubsec:ad_synergies}, where we give an example and compare how backwards elimination and Shapley values divide these “extra” conversions.

\subsubsection{Incremental Attribution}
\label{subsubsec:incremental_attribution}
Suppose that we have built a model for $\hat{\lambda}(t)$ with experimental data and separate ad and query effects. In this case we want to attribute credit to ads based only on the increase in conversion intensity caused by ads and not the query effect. For concreteness, suppose the model is the same as equation \ref{eq:incr_model}, i.e.:
\begin{align}
    \log(\lambda(t)) = \alpha_0 + \alpha_{\text{age bucket}} &+ \sum_j f(t - t_j) I\{\text{ad $j$ shown}\} + \sum_{j,k} g_k(t-t_j, x_{jk}) I\{\text{ad $j$ shown}\} \nonumber \\
& {} + \sum_j m(t - t_j) +\sum_{j,k} n_k(t-t_j, x_{jk})
\end{align}
We define the attribution credit in the same way as before, removing a single ad at a time and observing the change in $\hat{\lambda}(\tstar)$, but now we keep all the query effects throughout. Assuming, as before, that there are exactly $n$ ads before the conversion, we have
\begin{align}
    \log(\hat{\lambda}(\tstar, \mathcal{A}(l)) = \hat{\alpha}_0 + \hat{\alpha}_{\text{age bucket}} &+ \sum_{j=1}^l \hat{f}(t - t_j) I\{\text{ad $j$ shown}\} + \sum_{j=1}^l \sum_k \hat{g}_k(t-t_j, x_{jk}) I\{\text{ad $j$ shown}\} \nonumber \\
& {} + \sum_{j=1}^n \hat{m}(t - t_j) +\sum_{j=1}^n \sum_k \hat{n}_k(t-t_j, x_{jk})
\end{align}
That is, the summations for $\hat{f}$ and $\hat{g}_k$, the ad effects, consider only the ads in $\mathcal{A}(l)$, while the summations for $\hat{m}$ and $\hat{n}_k$, the query effects, consider all of the $n$ events preceding the conversion. The overall raw credit given to ads is therefore the difference between the estimated conversion intensity at $\tstar$ and the estimated conversion intensity for a counterfactual path with the same queries, but where ads were always withheld. As with the non-incremental credit, it can be shown that the expected value of the total normalized ad credit for each user equals the expected number of incremental conversions (i.e. the expected difference in conversions with and without ads) for that path. See the appendix for details. Thus while it may seem strange to have an ad’s credit depend on features of later queries, this actually leads to the most intuitive value of the individual and total ad credit.

\subsubsection{Ad Synergies and Comparison to Shapley Values}
\label{subsubsec:ad_synergies}
We mentioned previously that backwards elimination assigns credit for conversions requiring multiple ads to the last ad in the group. In this section, we make that more precise and compare how backwards elimination and Shapley values divide this credit. We start by defining what we mean by an ad’s marginal effect as well as “conversions requiring multiple ads.”

Define the marginal credit of a set of one or more ads $\mathcal{A}$ to be $m(\mathcal{A}) = \hat{\lambda}(\mathcal{A}) - \hat{\lambda}(\emptyset)$, where we have dropped the $\tstar$ for brevity. This marginal credit is exactly the difference in conversion intensity at conversion time with exactly these ads versus without any ads.  When our algorithm starts, it has $m(\mathcal{A}(n))$ units of raw credit to distribute and when it gets to the $j^{th}$ ad it has $m(\mathcal{A}(j))$ units of credit left to distribute, by definition. If we think of the algorithm as splitting credit between two groups of ads, $\mathcal{A}(j-1)$ and the singleton $j^{th}$ ad, $\{A_j\} = \{(t_j, X_j)\}$, then the credit given to the earlier ads, $\mathcal{A}(j-1)$, is exactly their marginal credit. Meanwhile, the credit assigned by the algorithm for the later ad, $\{A_j\}$, is $m(\mathcal{A}(j)) - m(\mathcal{A}(j-1))$. 

Then we can define the synergy due to a user seeing both $\mathcal{A}(j-1)$ and $\{A_j\}$ as 
\begin{equation}
    S(\mathcal{A}(j-1), A_j) = m(\mathcal{A}(j)) - m(\mathcal{A}(j-1)) - m(A_j)
\end{equation}
i.e. the difference between the credit received by $A_j$ when seen after the ads in $\mathcal{A}(j-1)$ and the marginal ad credit for a path containing only $\{A_j\}$\footnote{In fact, the synergy can be defined more generally for an ad $A$ and a set of ads $\mathcal{A}$ as $S(\mathcal{A}, \{A\}) = m(\mathcal{A} \cup {A}) - m(\mathcal{A}) - m(A)$, but we will not require this more general definition.}. When the conversion intensity is superadditive, so that showing multiple ads leads to more conversions than the sum of the marginal  increase in conversions from showing each ad individually,  $S(\mathcal{A}(j-1), A_j)$ is positive and backwards elimination distributes the extra synergy credit to the later ad. When the conversion intensity is subadditive, so that showing multiple ads leads to fewer conversions than the sum of the marginal increase in conversions from showing ad individually,  $S(\mathcal{A}(j-1), A_j)$ is negative, and the later ad gets credit smaller than its marginal credit.

\textbf{Example:} Suppose that a user path has 2 ads, $A_1, A_2$, prior to a conversion and suppose that $\hat{\lambda}(\tstar, \{A_1, A_2\}) = \exp(f_1(\tstar-t_1) + f_2(\tstar-t_2))$, i.e. with no other interactions and with the baseline rate equal to 1. Suppose also that $\hat{\lambda}(\tstar, \{A_1\}) = \exp(f_1(\tstar-t_1)) = 2$, $\hat{\lambda}(\tstar, \{A_2\}) = \exp(f_2(\tstar-t_2)) = 3$, in other words the contribution from each ad is the same as if it were the only ad in the path. Then the overall raw ad credit for the 2-ad path will be $m(\{A_1, A_2\}) = \hat{\lambda}(\tstar) - \hat{\lambda}(\emptyset) = 2*3 - 1 = 5$. The raw marginal contribution of these ads is $m(A_1) = 1$, $m(A_2) = 2$.  The raw credit of these ads is
\begin{align}
    RawCredit(A_2) &= 2*3 - 2 = 4 \\
    RawCredit(A_1) &= 2 - 1 = 1
\end{align}
We can see that the first ad receives credit equal to its marginal effect. The second ad receives credit equal to its marginal effect, ($m(A_2) = 2$), plus the synergy between the first and second ads, $S(\{A_1\}, A_2) = m(\{A_1, A_2\}) - m(A_1) - m(A_2) = 5 - 1 - 2 = 2$.

We make two additional remarks regarding this example:
\begin{enumerate}
    \item In general, the synergy between ads will decrease as the time between ads increases. More precisely, if we assume the parameterization in the example, some straightforward algebra shows that 
    \begin{equation}
        S(\{A_1\}, A_2) = \left(\exp(f_1(\tstar-t_1)) - 1\right) \left(\exp(f_2(\tstar-t_2)) - 1\right)
    \end{equation}
    Assuming the $f_i$ are decreasing functions, and fixing both the conversion time, $\tstar$, as well as the time between the last ad and the conversion, $\tstar-t_2$, then as the time between the ads, $t_2-t_1$, increases, the synergy will generally decrease, since $\tstar-t_1$ will increase, causing the first term to decrease. In other words, if the ads are far apart, their synergy will be smaller.
    \item While the synergy in the example is positive, negative synergy can occur when there is diminishing or even zero marginal benefit to showing multiple ads, perhaps due to user ad fatigue. As a numerical example, consider the case where $\hat{\lambda}(\tstar, \{A_1\}) = 2$ and $\hat{\lambda}(\tstar, \{A_2\}) = 3$ as before, but $\hat{\lambda}(\tstar, \{A_1, A_2\}) =3$. In other words, the fact that the user has seen multiple ads doesn’t increase the intensity; only the timing of the last ad matters. This can be parameterized by allowing a dependence between ads, as described in Equation \ref{eq:ad_interactions}. In this case $RawCredit(A_1) = 1$ as before, but $RawCredit(A_2) = 3 -2 = 1$, which is less than $m(A_2)$, implying that the synergy is negative and that we would be better off showing the ads to different users rather than the same user.
\end{enumerate}

Giving all of this synergy or interaction credit to the last ad may seem unfair, since the additional increase or decrease in the conversion intensity only happens if \textbf{all} of the ads occur. However, recall that one potential use case for attribution is as an input to bidding, leading to higher bids on ads that drive more conversions. Giving the synergy to the last ad rather than the earlier ad reflects our knowledge when we are bidding on the earlier ad. At that time, without either additional assumptions or modeling, we do not know if this user will have future ad impressions for us to bid on, therefore one reasonable approach is to bid based on the marginal effect of this current ad, together with its synergies with past ads, while ignoring any synergies between the ad being bid on and future ads (which may or may not occur). Thus, in this application, backwards elimination might be desirable precisely because it gives all the synergy credit to the last ad.

Other methodologies, such as the popular Shapley value method used in \cite{dalessandro2012causally,shao2011data,du2019causally}, split this synergy evenly amongst the ads involved. Shapley values come from game theory and try to fairly divide the payoff (increase in intensity) for a coalition (group of ads) amongst the players (individual ads). It does this by giving each player credit proportional to its average marginal contribution to all possible coalitions (subsets of ads). More precisely, in a game with player set $\Omega, (|\Omega|=N)$ and value function $v$, the payoff to player $i$ is given by:
\begin{equation}
    \phi_j(v) = \sum_{O \subseteq \Omega \setminus \{j\}} \frac{1}{N!}  |O|!  (N-|O|-1)!  \left[v(O \cup \{j\}) - v(O)\right]
\end{equation}
In the case of ad attribution, the value function is $v(\cdot)=\hat{\lambda}(\tstar, \cdot)$ (or equivalently, $v(\mathcal{A}) = \hat{\lambda}(\tstar, \mathcal{A}) - \hat{\lambda}(\tstar, \emptyset)$ for any set $\mathcal{A}$), while the set of players equals the set of all ads before the conversions, i.e. $\Omega = \mathcal{A}(n)$. $j$ is the index of the ad receiving credit.

For a path with only a single ad, both backwards elimination and Shapley values will result in the same credit assignment. Similarly, the total credit given to all ads (as opposed to the baseline), is also the same for both methods. However, if there are 2 or more ads with nonzero synergy, the credit assignment will differ.

\textbf{Example:} Continuing with our previous example, it is straightforward that the ad credit for $A_1$ will be
\begin{equation}
    ShapleyRawCredit(A_1) = \frac{1}{2} \left( \hat{\lambda}(\tstar, \{A_1\}) - \hat{\lambda}(\tstar, \emptyset)\right) + \frac{1}{2} \left( \hat{\lambda}(\tstar, \{A_1, A_2\}) - \hat{\lambda}(\tstar, \{A_2\})\right) = 2
\end{equation}
Similarly, we will have
\begin{equation}
    ShapleyRawCredit(A_2) = 3
\end{equation}
This gives more credit to the first ads compared to the Backwards Elimination method. In fact, its credit equals its marginal effect, 1, plus half of the synergy between the ads, $1/2 * 2$. Similarly, the credit for the second ad equals its marginal effect plus half the synergy.

More generally, it can be shown that for synergy requiring $k$ ads, Shapley values gives each ad additional credit equal to $1/k$ of that synergy. The proof is similar to Theorem 1 in \cite{zhao2018shapley}. For some applications or media types, this strategy of dividing the synergy may be preferred. For example, when retrospectively examining data for a group of ad campaigns splitting the synergy may be a more desirable way to compare the relative contributions of the campaigns.

While backwards elimination and Shapley values divide the synergy credit differently, both depend on an ad’s contribution to $\lambda(\tstar)$. In particular, this implies that given a suitable model for $\lambda(t)$, both methods will satisfy Requirement \#2 from Section \ref{subsec:req2}: conversion credit for an ad will decrease the further it is from conversion time, and the credit given to ads depends not just on their order, but on their spacing in time\footnote{Requirement \#1, that we handle incomplete or censored data, depends on only on $\lambda(t)$, not the attribution methodology.}. In that sense, both are reasonable choices. Given a model for $\lambda(t)$, the two methods can be compared based on the specific goals of the application.

\section{Evaluation}
\label{sec:eval}
To evaluate this system, we use standard model fit metrics for a Poisson regression, such as the log-likelihood or Poisson loss. There is also the prediction bias: predicted conversions / observed conversions - 1. We can also consider sliced versions of these metrics to aid in model comparison for feature selection.

With experimental data, we can estimate the ground truth number of incremental conversions that are caused by ads without needing a model by simply comparing the number of conversions in the exposed and unexposed groups. To evaluate our model, which is fit using both the exposed and unexposed users, we can compare this ground truth estimate to the predicted incremental conversions obtained by comparing the model’s predicted number of conversions in the exposed and unexposed groups. If our goal is to correctly model incrementality, comparing the ground truth and predicted versions of incremental conversions (i.e the difference between groups) may be of greater interest than comparing the actual and predicted conversions in each group. As a way of evaluating both our model and credit assignment methodology, we can also compare the ground truth incremental conversions to the normalized version of the ad credit. To justify this comparison, recall that as discussed in Section \ref{subsubsec:incremental_attribution}, the expected value of the ad credit should equal the expected number of incremental conversions, assuming the model estimates are correct.

The exact estimators for these metrics depends on the details of how the experiment is run, which in turn depend on the details of the media type and ad serving environment. For simplicity and specificity, suppose again that the experiment splits users for the duration of the experiment into either an exposed group, which sees ads as normal, or an unexposed group, for whom ads are withheld, but that we are able to observe the conversions of both groups. Then we can consider the following ground truth metric, which we call incremental conversions per user:
\begin{equation}
    ICPU = \frac{\text{conversions in exposed group}}{\text{users in exposed group}} - \frac{\text{conversions in unexposed group}}{\text{users in unexposed group}}
\end{equation}
This is essentially the observed difference in the average conversion rate between exposed and unexposed groups. A closely related metric is the rate of incremental conversions per unit time:
\begin{equation}
    ICPT = \frac{\text{conversions in exposed group}}{\text{total observation time for users in exposed group}} - \frac{\text{conversions in unexposed group}}{\text{total observation time for users in unexposed group}}
\end{equation}
where the total observation time is defined as
\begin{equation}
    \text{total observation time for users in } X \text{ group} = \sum_{i: i \in X} \text{total time that user } i \text{ is observed in the experiment}
\end{equation}
If all users are observed for the same length of time in the experiment, then ICPT differs from ICPU by a constant factor. However, if the average time that each user is observed for differs between exposed and unexposed, then the difference is more complicated and ICPT may be a more useful metric.

We can also consider ICPE or incremental conversions per exposed conversion:
\begin{equation}
    ICPE = ICPU \times \frac{\text{users in exposed group}}{\text{conversions in exposed group}}
\end{equation}
This represents the proportion of conversions in the exposed group that are incremental, after normalizing for any differences in the sizes of the exposed and unexposed groups. As with ICPT compared to ICPU, if the average time that each user is observed for differs between exposed and unexposed, we may prefer a version of ICPE that accounts for this:
\begin{equation}
    ICPE' = ICPT \times \frac{\text{total observation time for users in exposed group}}{\text{conversions in exposed group}}
\end{equation}

We can compare each of these with their counterparts predicted by the model, which we call PICPU (predicted incremental conversions per user) and PICPPE (predicted incremental conversions per predicted exposed conversion). Their definitions are the same, but with “conversions” replaced by “predicted conversions”. We can also evaluate our attribution methodology by considering the attribution credit, normalized by $\hat{\lambda}(\tstar)$, given to ads for conversions that occur in the exposed group. This metric, which we call AICPE (attributed incremental conversions per exposed conversion), represents the proportion of credit for exposed conversions that is given to ads and is comparable to PICPPE and ICPE. Comparing PICPPE and AICPE against ICPE is therefore an appropriate validation metric for our model and attribution methodology, respectively. Confidence intervals for these metrics can be obtained by bootstrapping over users. Alternatively, we could incorporate the uncertainty over the model itself by bootstrapping or using a block jackknife to refit the model and compute the evaluation metrics per block and then compute the variability of this estimate over the blocks.

For feature selection, it can be helpful to compare these metrics on various slices of users, however we must be cautious in choosing features to slice on. In particular, we must avoid any confounding between a user’s treatment assignment, slice value, and conversion outcome. For example, if ad exposure increases the length of a user’s path, perhaps because seeing ads leads to increased activity, then slicing by the total number of queries in a path leads to incomparable sets of exposed and unexposed users. However slicing on user features that are unaffected by the experiment, such as city or gender, is still valid and can be useful both for feature selection and understanding user behavior.

\section{Simulation Study}
\label{sec:sims}
We consider several simulation scenarios and show that our system performs well in these. In general, simulating user paths and conversion behavior requires strong assumptions about the generating process for both ad exposures and conversions. These assumptions are necessarily much simpler than actual user behavior. Nevertheless, simulations can still provide confidence by showing that the model is working as expected in these cases.

For each of the scenarios below, we simulate 500 distinct data sets, each with 1 million users and 30 days of data per user. We report the average coefficient estimates across these 500 data sets, and use the 0.025 and 0.975 quantiles across data sets to construct 95\% confidence intervals.

\subsection{Scenario 1}
Here we simulate exactly one ad per user, with the ad occurrence time being uniform in our 30 day observation window, i.e. $[0, 30.0]$. We simulate a pre-query conversion rate of 1 conversion / 30 days and an ad effect that doubles the conversion rate on the first day after the ad, increases it by a factor of 1.5 on the second day, and increases it by a factor of 1.2 after that. We refer to these as the short, medium, and long term effects, respectively. This scenario fits the model in Equation \ref{eq:curr_tedda_model}:
\begin{equation}
    \log(\lambda(t)) = \alpha_0 + \beta_1 I\{t-t_1 \leq 1\} + \beta_2 I\{1<t-t_1 \leq 2\} +
                       \beta_3 I\{2<t-t_1 \leq 30\}
\end{equation}
where we have rescaled time to be measured in days rather than hours.

Since we are assuming a piecewise constant ad effect, the conversion intensity is piecewise constant overall and so we can simulate from it by simulating from a Poisson distribution with appropriate mean and offset. We then use this data to fit the model above and repeat this for each of the 500 simulated datasets, resulting in 500 model estimates, each fit on an independent set of 1 million users. The table shows the true values used in our simulations, together with the parameter estimates (averaged over the 500 repetitions) and CI’s (from the observed quantiles across 500 repetitions).  

\begin{table}[h]
    \centering
    \begin{tabular}{|c|c|c|c|}
    \hline
         & Ground Truth & \makecell[c]{Mean\\(across 500 datasets)} & \makecell[c]{CI\\{[p2.5, p97.5]}\\(across 500 datasets)}  \\
    \hline
    \makecell[c]{Baseline (per day)\\$[\exp(\alpha_0)]$} & 0.0333 & 0.0333 & [0.0332, 0.0334] \\
    \hline
    \makecell[c]{Short term\\$[\exp(\beta_1)]$} & 2.0 & 2.000 & [1.983, 2.017] \\
    \hline
    \makecell[c]{Medium term\\$[\exp(\beta_2)]$} & 1.5 & 1.498 & [1.485, 1.583] \\
    \hline
    \makecell[c]{Long term\\$[\exp(\beta_3)]$} & 1.2 & 1.200 & [1.195, 1.204] \\
    \hline
    \end{tabular}
    \caption{Ground truth and estimated model coefficients for Scenario 1.}
\end{table}

Using our attribution methodology, and normalizing by $\lambda(\tstar)$, we get a mean AICPE of 11.94\% [11.76\%, 12.21\%], with the remaining credit going to the baseline. This model was not fit on simulated experimental data and no query effect was fit, so there is technically no ICPE to compare this to. However, we can compare it to what the ICPE would be if we assumed there was no query effect and we simulated additional users who have an ad query but don’t see an ad, i.e. whose conversion intensity is equal to the baseline for the entire observation window. This is equivalent to an experimental model where there is no query effect: i.e. when the observed and incremental effect of ads is the same. The corresponding ICPE for these users is 11.95\% [11.83\%, 12.06\%].

\subsection{Scenario 2}
In this scenario we introduce a second type of ad for which there is no long-term effect. In particular, we assume that this second type of ad increases the conversion intensity by a factor of 1.5x on the first day, 1.2x on the second day and 1.0x (or no increase) after that.

We will simulate each user as having exactly 2 ads, 1 of each type. The occurrence times for each ad are independent of each other, with each being uniform in $[0, 30]$. This assumption allows us to simplify the model notation to 
\begin{align}
    \log(\lambda(t)) = \alpha_0 &+ \beta_1 I\{0<t-t_{\text{ad type 1}} \leq 1\} + \beta_2 I\{1<t-t_{\text{ad type 1}} \leq 2\} + \beta_3 I\{2<t-t_{\text{ad type 1}} \leq 30\} \nonumber \\
    {} &+ \beta_4 I\{0<t-t_{\text{ad type 2}} \leq 1\} + \beta_5 I\{1<t-t_{\text{ad type 2}} \leq 2\} + \beta_6 I\{2<t-t_{\text{ad type 2}} \leq 30\}
\end{align}
where $t_{\text{ad type i}}$ is the occurrence time for the ad of type $i$. The table shows the true values used in our simulations, together with the parameter estimates and CIs, derived as before.

\begin{table}[h]
    \centering
    \begin{tabular}{|c|c|c|c|c|}
    \hline
    \multicolumn{2}{|c|}{} & Ground Truth & \makecell[c]{Mean\\(across 500 datasets)} & \makecell[c]{CI\\{[p2.5, p97.5]}\\(across 500 datasets)}  \\
    \hline
    \multicolumn{2}{|c|}{\makecell[c]{Baseline (per day)\\$[\exp(\alpha_0)]$}} & 0.0333 & 0.0333 & [0.0332, 0.0334] \\
    \hline
    \multirow{3}{*}{Ad Type 1} & \makecell[c]{Short term\\$[\exp(\beta_1)]$} & 2.0 & 2.000 & [1.983, 2.016] \\
    \cline{2-5}
    & \makecell[c]{Medium term\\$[\exp(\beta_2)]$} & 1.5 & 1.498 & [1.483, 1.512] \\
    \cline{2-5}
    & \makecell[c]{Long term\\$[\exp(\beta_3)]$} & 1.2 & 1.200 & [1.195, 1.204] \\
    \hline
    \multirow{3}{*}{Ad Type 2} & \makecell[c]{Short term\\$[\exp(\beta_4)]$} & 1.5 & 1.498 & [1.485, 1.511] \\
    \cline{2-5}
    & \makecell[c]{Medium term\\$[\exp(\beta_5)]$} & 1.2 & 1.199 & [1.188, 1.210] \\
    \cline{2-5}
    & \makecell[c]{Long term\\$[\exp(\beta_6)]$} & 1.0 & 1.000 & [1.000, 1.002] \\
    \hline
    \end{tabular}
    \caption{Ground truth and estimated model coefficients for Scenario 2.}
\end{table}

Using our attribution methodology, and normalizing by $\lambda(\tstar)$, we get a mean AICPE of 13.90\% [13.72\%, 14.09\%], with the remaining credit going to the baseline. As before, the model was not fit on simulated experimental data and no query effect was fit, so there is technically no ICPE to compare to. However, we can again compare to what the ICPE would be if we assumed there was no query effect and simulated additional users who have an ad query but don’t see an ad and whose conversion intensity is therefore equal to the baseline for the entire observation window. The corresponding ICPE is then 13.91\% [13.79\%, 14.02\%].

\subsection{Scenario 3}
In this scenario we simulate the second ad type as having no effect on user’s conversion rates. This could occur for public service announcements or similar types of ads. The other simulation details and model formulation are as in Scenario 2. The table below shows the true values used in our simulations, together with the parameter estimates and CI’s, derived as before.

\begin{table}[h]
    \centering
    \begin{tabular}{|c|c|c|c|c|}
    \hline
    \multicolumn{2}{|c|}{} & Ground Truth & \makecell[c]{Mean\\(across 500 datasets)} & \makecell[c]{CI\\{[p2.5, p97.5]}\\(across 500 datasets)}  \\
    \hline
    \multicolumn{2}{|c|}{\makecell[c]{Baseline (per day)\\$[\exp(\alpha_0)]$}} & 0.0333 & 0.0333 & [0.0332, 0.0334] \\
    \hline
    \multirow{3}{*}{Ad Type 1} & \makecell[c]{Short term\\$[\exp(\beta_1)]$} & 2.0 & 2.000 & [1.983, 2.017] \\
    \cline{2-5}
    & \makecell[c]{Medium term\\$[\exp(\beta_2)]$} & 1.5 & 1.498 & [1.483, 1.512] \\
    \cline{2-5}
    & \makecell[c]{Long term\\$[\exp(\beta_3)]$} & 1.2 & 1.200 & [1.195, 1.204] \\
    \hline
    \multirow{3}{*}{Ad Type 2} & \makecell[c]{Short term\\$[\exp(\beta_4)]$} & 1.0 & 1.000 & [0.992, 1.007] \\
    \cline{2-5}
    & \makecell[c]{Medium term\\$[\exp(\beta_5)]$} & 1.0 & 1.000 & [0.993, 1.007] \\
    \cline{2-5}
    & \makecell[c]{Long term\\$[\exp(\beta_6)]$} & 1.0 & 1.000 & [1.000, 1.003] \\
    \hline
    \end{tabular}
    \caption{Ground truth and estimated model coefficients for Scenario 3.}
\end{table}

Using our attribution methodology, and normalizing by $\lambda(\tstar)$, we get a mean AICPE of 11.94\% [11.73\%, 12.12\%], with the remaining credit going to the baseline. This is nearly the same as in Scenario 1, which is expected since the second ad type has no effect. As before, the model was not fit on simulated experimental data, so there is technically no ICPE to compare to, but we can compare to what the ICPE would be if we assumed there was no query effect and simulated additional users who have an ad query but don’t see an ad. The corresponding ICPE is then 11.95\% [11.84\%, 12.07\%].

\subsection{Scenario 4}
In this scenario we use the same two types of ads as in Scenario 2, but allow the number of ads per user to vary. We also allow more freedom in the model we fit than necessary to describe the data generating process for the simulations. The details are below.

We allow the number of ads per user to be either 1, 2, or 3. The probability of each is proportional to the probability of the corresponding number of events for Poisson random variable with mean 2. Equivalently, we can think of the events for each user as being a Poisson(2) random variable that is clipped to be between 1 and 3. On average, 40.6\% of the users have a single ad in the path, 27.1\% have 2 ads, and 32.3\% have 3 ads. Each ad is equally likely to be of either type. So conditional on a user seeing 3 ads total, they may see 0, 1, 2 or 3 ads of each type, as long as the overall number of ads is 3. The ad occurrence times are still independent and uniform on our 30 day observation window, i.e. on [0, 30].

Recall that ad type 1 increased the conversion intensity by a factor of 2x on the first day, 1.5x on the second day, and 1.2x after that, while ad type 2 increased the conversion intensity by a factor of 1.5x on the first day, 1.2x on the second day and 1.0x (no increase) after that. We assume that the effect of a second or third ad of the same type is the same as the effect of the first ad. For example, if a user sees an ad of type 1 at time 0 and a second ad of type 1 one hour later, then the user’s ad intensity will be 2*2 = 4x higher than if they had seen no ads at all. More generally, we simulate the intensity as
\begin{align}
    \log(\lambda(t)) = \alpha_0 &+ \beta_1 \#\{i:0<t-t_i\leq 1, \text{ad type}(i) = 1\} + \beta_2 \#\{i:1<t-t_i \leq 2, \text{ad type}(i) =1\} \nonumber \\
    &\qquad \qquad {} + \beta_3 \#\{i:2<t-t_i \leq 30, \text{ad type}(i) = 1\} \nonumber \\
    &+ \beta_4 \#\{i:0<t-t_i\leq 1, \text{ad type}(i) = 2\} + \beta_5 \#\{i:1<t-t_i \leq 2, \text{ad type}(i) =2\} \nonumber \\
    &\qquad \qquad {} + \beta_6 \#\{i:2<t-t_i \leq 30, \text{ad type}(i) = 2\}
\end{align}
where $\#\{i: conditions\}$ denotes the number of ads $i$ satisfying the condition. Equivalently we could write
\begin{align}
    \log(\lambda(t)) = \alpha_0 &+ \sum_{i: \text{ad type}(i) = 1} \big[ \beta_1 I\{0<t-t_i \leq 1\} + \beta_2 I\{1<t-t_i \leq 2\} + \beta_3 I\{2<t-t_i \leq 30\} \big] \nonumber \\
    {} &+ \sum_{i: \text{ad type}(i) = 2} \big[ \beta_4 I\{0<t-t_i \leq 1\} + \beta_5 I\{1<t-t_i \leq 2\} + \beta_6 I\{2<t-t_i \leq 30\} \big]
\end{align}

However, when we fit the model, we do not assume that the second and third ad of the same type have the same effect as the first. Instead we allow for more degrees of freedom. For this section we will use $\gamma$ to denote the coefficients estimated by the model, while continuing to use $\beta$ for the true coefficients used to generate the simulation data. Then the model we fit is
\begin{align}
    \log(\lambda(t)) = \alpha_0 + \sum_{k=1}^3 \Big[ &\gamma_{11k} I\{\text{exactly $k$ type 1 ads with } 0<t-t_i \leq 1\}  \nonumber \\
    {} + & \gamma_{12k} I\{\text{exactly $k$ type 1 ads with } 1<t-t_i \leq 2\} \nonumber \\
    {} + & \gamma_{13k} I\{\text{exactly $k$ type 1 ads with } 2<t-t_i \leq 30\} \Big] \nonumber \\
    {} + \sum_{k=1}^3 \Big[ &\gamma_{21k} I\{\text{exactly $k$ type 2 ads with } 0<t-t_i \leq 1\}  \nonumber \\
    {} + & \gamma_{22k} I\{\text{exactly $k$ type 2 ads with } 1<t-t_i \leq 2\} \nonumber \\
    {} + & \gamma_{23k} I\{\text{exactly $k$ type 2 ads with } 2<t-t_i \leq 30\} \Big]
\end{align}
For each $\gamma_{ijk}$, $i$ indexes the ad type (1 or 2), $j$ indexes the time interval or term (1 for (0, 1] (short term), 2 for (1, 2] (medium term), and 3 for (2, 30] (long term)), while $k$ indexes the exact number of ads of type $i$ in interval type $j$ that is under consideration. In particular, it is straightforward to see that the ground truth value for $\gamma_{11k}$ is $k*\beta_1$ or on the original scale $\exp(\gamma_{11k})$ should equal $\exp(k*\beta_1)$. Similarly, $\gamma_{12k}$ corresponds to $k*\beta_2$ or on the original scale $\exp(\gamma_{12k})$ corresponds to $\exp(k*\beta_2)$.

As before, we simulate 500 independent data sets, each with 1 million users, and fit models independently on each data set. The table below shows the average parameter estimate across the 500 data sets as well as the CI, derived by taking the 2.5\% and 97.5\% quantiles of the estimates across the 500 data sets. The first three columns specify the parameter being estimated, corresponding to $i$, $j$, and $k$ in $\gamma_{ijk}$. The fourth column gives the ground truth, giving both the numerical value and the formulation in terms of the $\beta_i$. The last two columns are the mean and CI.

\begin{table}[htp]
    \centering
    \begin{tabular}{|c|c|c|c|c|c|}
    \hline
    \multicolumn{3}{|c|}{Parameter to Estimate} & & & \\
    \hline
    Ad type & Time Term & \# ads in term & Ground Truth & \makecell[c]{Mean\\(across 500 datasets)} & \makecell[c]{CI\\{[p2.5, p97.5]}\\(across 500 datasets)}  \\
    \hline
    \multicolumn{3}{|c|}{\makecell[c]{Baseline (per day)\\$[\exp(\alpha_0)]$}} & 0.0333 & 0.0333 & [0.0332, 0.0334] \\
    \hline
    \multirow{9}{*}{Ad Type 1} & \multirow{3}{*}{Short Term} & \makecell[c]{1 ad\\$[\exp(\gamma_{111})]$} & \makecell[c]{2.0\\$[\exp(\beta_1)]$} & 1.999 & [1.983, 2.015] \\
    \cline{3-6}
    & & \makecell[c]{2 ads\\$[\exp(\gamma_{112})]$} & \makecell[c]{4.0\\$[\exp(2\beta_1)]$} & 3.984 & [3.784, 4.183] \\
    \cline{3-6}
    & & \makecell[c]{3 ads\\$[\exp(\gamma_{113})]$} & \makecell[c]{8.0\\$[\exp(3\beta_1)]$} & 4.906 & [2.523, 7.746] \\
    \cline{2-6}
     & \multirow{3}{*}{Medium Term} & \makecell[c]{1 ad\\$[\exp(\gamma_{121})]$} & \makecell[c]{1.5\\$[\exp(\beta_2)]$} & 1.498 & [1.485, 1.512] \\
    \cline{3-6}
    & & \makecell[c]{2 ads\\$[\exp(\gamma_{122})]$} & \makecell[c]{2.25\\$[\exp(2\beta_2)]$} & 2.213 & [2.059, 2.386] \\
    \cline{3-6}
    & & \makecell[c]{3 ads\\$[\exp(\gamma_{123})]$} & \makecell[c]{3.375\\$[\exp(3\beta_2)]$} & 1.850 & [0.666, 3.676] \\
    \cline{2-6}
     & \multirow{3}{*}{Long Term} & \makecell[c]{1 ad\\$[\exp(\gamma_{131})]$} & \makecell[c]{1.2\\$[\exp(\beta_3)]$} & 1.200 & [1.196, 1.205] \\
    \cline{3-6}
    & & \makecell[c]{2 ads\\$[\exp(\gamma_{132})]$} & \makecell[c]{1.44\\$[\exp(2\beta_3)]$} & 1.438 & [1.429, 1.449] \\
    \cline{3-6}
    & & \makecell[c]{3 ads\\$[\exp(\gamma_{133})]$} & \makecell[c]{1.728\\$[\exp(3\beta_3)]$} & 1.724 & [1.692, 1.755] \\
    \hline

    \multirow{9}{*}{Ad Type 2} & \multirow{3}{*}{Short Term} & \makecell[c]{1 ad\\$[\exp(\gamma_{211})]$} & \makecell[c]{1.5\\$[\exp(\beta_4)]$} & 1.502 & [1.488, 1.516] \\
    \cline{3-6}
    & & \makecell[c]{2 ads\\$[\exp(\gamma_{212})]$} & \makecell[c]{2.25\\$[\exp(2\beta_4)]$} & 2.270 & [2.109, 2.430] \\
    \cline{3-6}
    & & \makecell[c]{3 ads\\$[\exp(\gamma_{213})]$} & \makecell[c]{3.375\\$[\exp(3\beta_4)]$} & 2.138 & [0.735, 4.346] \\
    \cline{2-6}
     & \multirow{3}{*}{Medium Term} & \makecell[c]{1 ad\\$[\exp(\gamma_{221})]$} & \makecell[c]{1.2\\$[\exp(\beta_5)]$} & 1.201 & [1.189, 1.211] \\
    \cline{3-6}
    & & \makecell[c]{2 ads\\$[\exp(\gamma_{222})]$} & \makecell[c]{1.44\\$[\exp(2\beta_5)]$} & 1.427 & [1.300, 1.548] \\
    \cline{3-6}
    & & \makecell[c]{3 ads\\$[\exp(\gamma_{223})]$} & \makecell[c]{1.728\\$[\exp(3\beta_5)]$} & 1.130 & [0.876, 1.958] \\
    \cline{2-6}
     & \multirow{3}{*}{Long Term} & \makecell[c]{1 ad\\$[\exp(\gamma_{231})]$} & \makecell[c]{1.0\\$[\exp(\beta_6)]$} & 1.000 & [0.998, 1.003] \\
    \cline{3-6}
    & & \makecell[c]{2 ads\\$[\exp(\gamma_{232})]$} & \makecell[c]{1.0\\$[\exp(2\beta_6)]$} & 1.000 & [0.998, 1.001] \\
    \cline{3-6}
    & & \makecell[c]{3 ads\\$[\exp(\gamma_{233})]$} & \makecell[c]{1.0\\$[\exp(3\beta_6)]$} & 1.000 & [0.998, 1.000] \\
    \hline
    \end{tabular}
    \caption{Ground truth and estimated model coefficients for Scenario 4.}
\end{table}

We notice that
\begin{itemize}
    \item For the short and medium term effects in the 3 ad case (3 ads in a term), the average parameter estimate can be quite far from the truth, and the confidence interval can be quite wide, especially relative to previous simulations.
    \item For the short and medium term effects in the 2 ad case, the average parameter estimate is reasonably accurate, but the CI’s are wider than before.
    \item For the long term effects in the 2 and 3 ad cases, the average parameter estimates are reasonably accurate and the CIs are only slightly wider than before.
    \item For short, medium, and long term effects, performance in the 1 ad case is comparable to previous simulations.
\end{itemize}

These results can be explained by the relative sums of the interval lengths for which the ad effects are active. In the remainder of this section we will give numerical results from the simulation as well as intuition. Readers who are not interested in the details may skip to the next section.

 The model fit in this scenario does not assume that the effect of the second or third ad is the same as the first. As a result, each $\gamma_{ijk}$ is estimated based only on data from the sections of user paths where it is active (i.e. where the corresponding indicator function is not 0). The amount of data can be quantified by summing up the interval lengths (i.e. summing the offsets where the feature is active). The table below gives the average (across the 500 data sets) interval length, rounded to the nearest day, that each $\gamma_{ijk}$ is active. Since the two ad types are symmetric in this respect, their results are quite similar - we show them side by side to save space.

\begin{table}[h]
    \centering
    \begin{tabular}{|c|c|c|c|}
    \hline
    \multicolumn{2}{|c|}{Parameter to Estimate} & \multirow{2}{*}{\makecell[c]{Ad Type 1\\Offsets (days)}} & \multirow{2}{*}{\makecell[c]{Ad Type 2\\Offsets (days)}} \\
    \cline{1-2}
    Time Term & \# of ads in term & & \\
    \hline
    
    \multirow{3}{*}{Short Term} & \makecell[c]{1 ad\\$[\exp(\gamma_{i11})]$} & 922,567 & 922,528 \\
    \cline{2-4}
    & \makecell[c]{2 ads\\$[\exp(\gamma_{i12})]$} & 9,972 & 9,977 \\
    \cline{2-4}
    & \makecell[c]{3 ads\\$[\exp(\gamma_{i13})]$} & 44 & 44 \\
    \hline
    
    \multirow{3}{*}{Medium Term} & \makecell[c]{1 ad\\$[\exp(\gamma_{i21})]$} & 891,290 & 891,257 \\
    \cline{2-4}
    & \makecell[c]{2 ads\\$[\exp(\gamma_{i22})]$} & 9,633 & 9,636 \\
    \cline{2-4}
    & \makecell[c]{3 ads\\$[\exp(\gamma_{i23})]$} & 43 & 42 \\
    \hline
    
    \multirow{3}{*}{Long Term} & \makecell[c]{1 ad\\$[\exp(\gamma_{i31})]$} & 8,173,729 & 8,172,853 \\
    \cline{2-4}
    & \makecell[c]{2 ads\\$[\exp(\gamma_{i32})]$} & 1,831,242 & 1,831,566 \\
    \cline{2-4}
    & \makecell[c]{3 ads\\$[\exp(\gamma_{i33})]$} & 229,930 & 230,007 \\
    \hline
    \end{tabular}
    \caption{Average observed offset length for model coefficients in Scenario 4.}
    \label{table:interval_lengths}
\end{table}
The interval lengths are in line with the observations we made earlier. Intuitively, we can think of these interval lengths as depending on two factors: frequency (number of users) for which $\gamma_{ijk}$ is active and the length of time per user that it is active.

First consider the frequency, or the number of users for which $\gamma_{ijk}$ is active. First recall that users have a 32\% chance of having 3 ads and a 27\% chance of having 2 ads total. Moreover, given that users have 3 (resp 2) ads total, the chance that they are all of the same type is only 1/4 (resp 1/2). Consider first the long-term effects of having multiple ads of the same type. We focus on the effect of having 3 ads of the same type: it is straightforward to see that this is the rarest of the three, and the computations for the case of 2 ads of the same type are more involved, since they must consider paths with both 2 ads total and 3 ads total. For the 3 ad case, we might expect all of the users with 3 ads of the same type (or ~ 1 million * 0.32 * 1/4  = 80K users) to eventually have the long-term effect of 3 ads of the same type be active. In fact, the average number of users with either  $\gamma_{133}$ or $\gamma_{233}$ active is ~65K, since we must account for the cases where the third ad occurs in the last two days of the observation period, in which case there is no point in the user path where the long-term effect of three ads is active. Nevertheless, the take-away is that a fairly large number of users will at some point in their path have active the long term effects of having multiple ads.

By contrast, the frequency of short and medium term effects is much lower. Not only does the user need to have 2 or 3 ads of the same type, but they must occur within a 24 hour period. This turns out to be fairly rare: on average, out of 1 million users there are approximately 262 users with 3 ads of one type within 24 hours and approximately ~80K users with 2 ads of one type within 24 hours of each other, again split evenly amongst the two types\footnote{In general, users with 3 ads of the same type within 24 hours will also have a brief window where the short or medium term effect of having 2 ads of the same type will also be active. However, at 262 such users, this does not affect the frequency much}.

Now consider the length of time for which the ad effects are active. The short and medium term effects of ads, regardless of the number of ads, can last at most 24 hours per ad, by definition. Depending on the number of ads in a user’s path, they can have at most 3 x 24hour intervals where the short or medium term effects of a single ad are active or 1 x 24 hour interval where the short or medium term effects of 2 or 3 ads are active. However in the 2 (resp 3) ad cases, the interval will generally be shorter than 24 hours, since for e.g. the short term, it will start at the time of the second (resp third) ad and end 24 hours after the first ad occurred\footnote{Unless a third ad of the same type occurs within 24 hours of the first, in which case the corresponding interval for the effect of two ads of the same type will be shorter.
}. The effect of having multiple ads in the medium term will similarly be based on the same interval but shifted forward by 24 hours - although it can be shorter if it is cut off by the end of the 30 day observation window. By contrast, the intervals for the long term effects can be much longer, up to 28 days long, with the length depending on the position of the ads in the path. While we still expect intervals for the long-term effects of 2 or 3 ads to potentially be shorter than those for just 1 ad, the intervals overall are generally much longer than for short and medium term effects. So all else equal, we should expect to have more data for estimating the long term effects of ads than for short or medium term effects, while also expecting the intervals for the effect of multiple ads to be shorter than those for a single ad.

While these explanations are heuristic, this intuition matches the results shown in Table \ref{table:interval_lengths} and can in principle be made more precise. They also give a sense of the amount of data needed to estimate the different parameters in the model. While this also depends on the model fitting procedure used (e.g. how much, if any, regularization is applied), we can see how the accuracy of the estimates change with the amount of data we have for a parameter.

For consistency with the previous sections, we also report the AICPE and ICPE. Using our attribution methodology, and normalizing by $\lambda(\tstar)$, we get a mean AICPE of 13.87\% [13.71\%, 14.02\%], with the remaining credit going to the baseline. While this is similar to Scenario 2, this is mostly a numerical coincidence: for any particular user, they may have more or fewer ads of each type relative to Scenario 2, and so the attributed ad credit for a user could go in either direction. It just happens to average out to a similar value. As before, the model was not fit on simulated experimental data, so there is technically no ICPE to compare to, but we can compare to what the ICPE would be if we assumed there was no query effect and simulated additional users who have an ad query but don’t see an ad. The corresponding ICPE is then 13.87\% [13.75\%, 13.99\%]. We can see that despite the fact that we have trouble estimating some of the rarer coefficients, the overall ad credit is still quite close to the ICPE, i.e. its accuracy is not much worse on average.

\section{Discussion}
\label{sec:discussion}
We have presented a data-driven attribution system based on estimating the effect of ads on a user’s conversion rate per unit of time. This system satisfies our previously outlined requirements, namely that it can handle incomplete or censored data as well as take into account the times as which ads occur, not just their order, when assigning credit. These features of our system make it appropriate for use in real-time bidding, although the details are beyond the scope of this paper given the many application-specific considerations. We have also discussed some examples of how to use covariates to model the conversion intensity over time, although again the detailed choice of covariates is highly application-dependent. In this section, we will discuss two areas of potential future work. The first relates to the relationship between ads, while the second concerns the effect of modeling assumptions on attribution outcomes.

Thus far, we have assumed that ads are independent of each other, in addition to the standard Poisson process assumption that conversions are independent of each other. However, for some types of media seeing an ad can lead to a user seeing more ads in the future. This can be a direct result of a user’s interactions with an ad, such a click on a display ad that then leads to the user seeing further related ads. It can also occur as a result of more indirect interactions, such as when seeing an ad prompts a user to search for a related term, leading to them seeing additional related search ads. In these cases, to the extent that the later ads are caused by the earlier ads, some of the credit for the later ads should arguably be redistributed to the earlier ads. Our system does not currently account for these effects when allocating credit. One way to remedy this might be by having a multi-stage model, where we first use all the ads to predict conversions. Consider the earlier example where display ads can cause search ads. Suppose that in our original model, the search ad gets 0.6 credit and the display ad gets 0.4. We can fit a second model where we use display ads as the events and search ads (rather than conversions) as the response. We can then allocate credit for the search ad between the preceding display ads and a baseline. If, for example, in this second model 30\% of the credit for the search ad occurring goes to the display ad, with the remaining 70\% going to the baseline, then we can reallocate 30\% of the conversion credit attributed to the search ad in the first model to the display ad, keeping 70\% of the credit with the search ad. As a result, the total credit for the display ad would be $0.4 + 0.3 * 0.6 = 0.58$ and the credit for the search ad would be $0.7 * 0.6 = 0.42$. In practice however, this requires a rich dataset in order to be able to detect these interactions and thus may not always be possible.

While we have discussed a few examples of features and structure for modeling $\lambda(t)$, we have not discussed in detail how the model structure influences attribution results. Consider the simplistic model from our first simulation, generalized for the case where there are multiple ads in the path:
\begin{align}
    \log(\lambda(t)) = \alpha_0 &+ \beta_1 I\{0 < t-t_1 \leq 1 \text{ for some i}\} \nonumber \\
                        {} &+ \beta_2 I\{1<t-t_1 \leq 2 \text{ for some i}\} \nonumber \\
                        {} &+ \beta_3 I\{2<t-t_1 \leq 30 \text{ for some i}\}
\end{align}
If there were two ads in the 24 hours before a conversion, the second ad would not change the estimated conversion intensity at conversion time and would therefore get 0 credit, i.e. if we assume there are only two ads in the path, $\hat{\lambda}(\tstar, \mathcal{A}(2)) - \hat{\lambda}(\tstar, \mathcal{A}(1)) = 0$. If we find it unrealistic to distribute all the credit to the first ad rather than the second, then in this particular case we can simply model the marginal effects of an additional ad in each time bucket, similar to the model in simulation scenario 4, except that rather than modeling the effect of $k$ ads, we model the change in the intensity when we go from $k-1$ to $k$ ads:
\begin{align}
    \log(\lambda(t)) = \alpha_0 &+ \beta_1 I\{0<t-t_i \leq 1\} + \beta_2 I\{1<t-t_i \leq 2\} + \beta_3 I\{0<t-t_i \leq 30 \} \nonumber \\
    {} &+ \beta_4 I\{0<t-t_i \leq 1 \text{ for two distinct values of i}\} \nonumber \\
    {} &+ \beta_5 I\{1<t-t_i \leq 2 \text{ for two distinct values of i}\} \nonumber \\
    {} &+ \beta_6 I\{0<t-t_i \leq 30 \text{ for two distinct values of i} \}
\end{align}
More generally, we can consider a model where we estimate a separate ad decay function for each ad as in Equation \ref{eq:model_ads_individually}:
\begin{equation}
    \log(\lambda(t)) = \alpha_0 + \sum_j f_j(t - t_j)
\end{equation}
However, we cannot model an infinite number of ads in each term. At some point either data sparsity or regularization will lead to some of the later ad effects being estimated as 0. As a result, in some cases the last ads before a conversion may get 0 credit. One alternative to avoid this is to pool data and estimate some ads as having the same effect. This can be done by assuming that all ads $j>j'$ have the same effect:
\begin{equation}
    \log(\lambda(t)) = \alpha_0 + \sum_{j=1}^{j'} f_j(t - t_j) + \sum_{j = j'}^J f_{j'}(t-t_j) 
\end{equation}
Or we could pool in a more sophisticated way, by e.g. supposing that an ad with no other ads in the preceding X days “resets” the counter and has the same effect as the very first ad. This pooling makes it more likely that the last ads before a conversion will get non-zero attribution credit.

The larger point here is that the model structure and the way in which we pool data to estimate the different ad decay curves has repercussions on the results of the attribution algorithm. We can of course use overall model fit metrics to guide our choices, but this points to the necessity of considering a wide range of models. Care may also be needed, since attribution results may differ most for long paths, where we most care about MTA, while overall model fit may be comparable, if long paths are relatively rare. As with most modeling choices, the right approach depends on the context and requires careful consideration from the modeller.

\section*{Acknowledgements}
The authors would like to thank both past and present members of the TEDDA team at Google for their contributions to developing this methodology, as well as members from other Ads teams for their tremendous support.

\bibliographystyle{plain}
\bibliography{refs}

\appendix
\section{Appendix: Expected Value of $NormalizedCredit(j)$}
\label{sec:appendix}
In section \ref{subsubsec:backwards_elimination} we claim that assuming that our estimated intensity is correct, we have
\begin{equation}
    E[NormalizedCredit(j)] = \int \hat{\lambda}(t, \mathcal{A}(j)) - \hat{\lambda}(t, \mathcal{A}(j-1))  \mathrm{d}t
\end{equation}
where the expectation is over all conversion occurrences in the path. Recall that
\begin{equation}
    NormalizedCredit(j) = \frac{\hat{\lambda}(\tstar, \mathcal{A}(j)) - \hat{\lambda}(\tstar, \mathcal{A}(j-1))}{\hat{\lambda}(\tstar, \mathcal{A}(n))}
\end{equation}
where without loss of generality, $n$ was taken to be the number of ads at or before time $\tstar$. Then we have
\begin{align}
    E\left[NormalizedCredit(j)\right] &= E\left[\frac{\hat{\lambda}(\tstar, \mathcal{A}(j)) - \hat{\lambda}(\tstar, \mathcal{A}(j-1))}{\hat{\lambda}(\tstar, \mathcal{A}(n))}\right] \nonumber \\
    &= \int_0^\infty \frac{\lambda(t, \mathcal{A}(j)) - \lambda(t, \mathcal{A}(j-1))}{\lambda(t)} \mathrm{d}Y(t) \\
\shortintertext{Where $Y(t)$ is the measure for the occurrences in the Poisson process (i.e. the conversion occurrences). Then we have}
&= \int_0^\infty \frac{\lambda(t, \mathcal{A}(j)) - \lambda(t, \mathcal{A}(j-1))}{\lambda(t)} \lambda(t) \mathrm{d}t \\
&= \int_{t_j}^\infty \lambda(t, \mathcal{A}(j)) - \lambda(t, \mathcal{A}(j-1)) \mathrm{d}t
\end{align}
as claimed. The last equality follows because the intensities of a path with the first $j-1$ ads only differs from the intensity of a path with the first $j$ ads after the $j^{th}$ ad has occurred.

For non-incrementality models, the integral above can be computed (replacing $\lambda$ with its estimate $\hat{\lambda}$) as long as the features and times of the first $j$ ads are known, since $\lambda(t, \mathcal{A}(j))$ does not depend on events after $t_j$. However, as discussed in Section \ref{subsubsec:incremental_attribution}, for incrementality models, $\lambda(t, \mathcal{A}(j))$ can depend on all query (i.e. non-ad) events between $t_j$ and $t$, in addition to all ad events up until $t_j$. Therefore the integral can only be computed if all query events (but not necessarily ad events) that ever occur are known. More practically, for $t$ much larger than $t_j$, we would expect the contribution from the $j^{th}$ ad to be effectively 0. In practice, it should suffice to assume that query events are known as long as $\lambda(t, \mathcal{A}(j)) - \lambda(t, \mathcal{A}(j-1))$ is non-negligible, which may be only a few days, depending on the data.

\end{document}